\documentclass{aa}  
\usepackage{graphicx}
\usepackage{txfonts}

\usepackage{multirow}
\usepackage{threeparttable}
\usepackage{hyperref}
\usepackage{placeins}
\hypersetup{colorlinks,citecolor=blue,filecolor=blue,linkcolor=blue,urlcolor=blue}

\begin{document} 

   \title{ALMA observations of CH$_{3}$COCH$_{3}$ and the related species CH$_{3}$CHO, CH$_{3}$OH, and C$_{2}$H$_{5}$CN in line-rich molecular cores}
   
   \titlerunning{CH$_{3}$COCH$_{3}$}

   \author{Chuanshou Li\inst{1}
          \and
          Sheng-Li Qin\inst{1}
          \and
          Tie Liu\inst{2}
          \and
          Xunchuan Liu\inst{2}
          \and
          Xiaohu Li\inst{3}
          \and
          Li Chen\inst{1}
          \and
          Hong-Li Liu\inst{1}
          \and
          Fengwei Xu\inst{4,5}
          \and
          Meizhu Liu\inst{6}
          \and
           Mengyao Tang\inst{7}
          \and
          Hongqiong Shi\inst{1}
          \and
          Tianwei Zhang\inst{8,9}
          \and
          Yuefang Wu\inst{5}
          }

   \institute{School of Physics and Astronomy, Yunnan University, Kunming 650091, People's Republic of China\\
             \email{\url{lichuanshou2021@163.com}, \url{qin@ynu.edu.cn}}
         \and
             Shanghai Astronomical Observatory, Chinese Academy of Sciences, 80 Nandan Road, Shanghai 200030, People's Republic of China\\
             \email{\url{liutie@shao.ac.cn}}
         \and
             Xinjiang Astronomical Observatory, Chinese Academy of Sciences, Urumqi 831399, China
         \and
             Kavli Institute for Astronomy and Astrophysics, Peking University, Beĳing, 100871, People's Republic of China
         \and
             Department of Astronomy, School of Physics, Peking University, Beĳing, 100871, People's Republic of China
         \and
             Center for Astrophysics, Guangzhou University, Guangzhou 510006, People's Republic of China
         \and
             Institute of Astrophysics, School of Physics and Electronic Science, Chuxiong Normal University, Chuxiong 675000, People's Republic of China
         \and
             I. Physikalisches Institut, Universit{\"a}t zu K{\"o}ln, Z{\"u}lpicher Stra{\ss}e 77, 50937 K{\"o}ln, Germany
         \and
             Research Center for Intelligent Computing Platforms, Zhejiang Laboratory, Hangzhou 311100, P.R.China
             }
 
  \abstract
  {Acetone (CH$_{3}$COCH$_{3}$) is a carbonyl-bearing complex organic molecule, yet interstellar observations of acetone remain limited. Studying the formation and distribution of CH$_{3}$COCH$_{3}$ in the interstellar medium can provide valuable insights into prebiotic chemistry and the evolution of interstellar molecules.}
  {We explore the spatial distribution of CH$_{3}$COCH$_{3}$ and its correlation with the O-bearing molecules acetaldehyde (CH$_{3}$CHO) and methanol (CH$_{3}$OH), as well as the N-bearing molecule ethyl cyanide (C$_{2}$H$_{5}$CN), in massive protostellar clumps.}
  {We observed 11 massive protostellar clumps using ALMA at 345 GHz, with an angular resolution of 0.7$^{\prime\prime}$$-$1.0$^{\prime\prime}$. Spectral line transitions were identified using the eXtended CASA Line Analysis Software Suite. We constructed integrated intensity maps of CH$_{3}$COCH$_{3}$, CH$_{3}$CHO, CH$_{3}$OH, and C$_{2}$H$_{5}$CN and derived their rotation temperatures, column densities, and abundances under the assumption of local thermodynamic equilibrium.}
  {CH$_{3}$COCH$_{3}$ is detected in 16 line-rich cores from 9 massive protostellar clumps: 12 high-mass cores (core mass $\textgreater$8 M$_{\bigodot}$), 3 intermediate-mass cores (2$-$8 M$_{\bigodot}$), and 1 low-mass core ($\textless$2 M$_{\bigodot}$). CH$_{3}$CHO and CH$_{3}$OH are also detected in all 16 cores, while C$_{2}$H$_{5}$CN is detected in 15. The integrated intensity maps reveal similar spatial distributions for CH$_{3}$COCH$_{3}$, CH$_{3}$CHO, CH$_{3}$OH, and C$_{2}$H$_{5}$CN. The line emission peaks of all four molecules coincide with the continuum emission peaks in regions without ultra-compact H{\sc ii} regions. Significant correlations are observed in the abundances of these molecules, which also exhibit similar average temperatures.}  
  {Our observational results, supported by chemical models, suggest that CH$_{3}$COCH$_{3}$, CH$_{3}$CHO, and CH$_{3}$OH originate from the same gas. The observed temperatures and abundances of CH$_{3}$COCH$_{3}$ are consistent with model predictions involving grain surface chemistry.}
  

   \maketitle

\section{Introduction} \label{sec:Introduction}
Carbonyl-bearing complex organic molecules (COMs) are considered potential precursors of complex biomolecules \citep{2007ApJ...661..899B, 2008OLEB...38..489J, 2011ApJ...743...60H, 2018ApJ...862...84F}. The formation mechanisms of carbonyl-bearing COMs have garnered significant interest in astrochemistry and astrobiology \citep{2007ApJ...661..899B, 2007ApJ...660.1588B, 2018ApJ...862...84F}. Acetone (CH$_{3}$COCH$_{3}$) is a representative carbonyl-bearing COM; it is characterized by its symmetric structure centered on the C=O group. It was the first COM containing ten atoms discovered in the interstellar medium \citep[ISM;][]{1987A&A...180L..13C, 2002ApJ...578..245S, 2014MNRAS.444.3792A}. Recent advances in radio telescope technology have enabled the detection of CH$_{3}$COCH$_{3}$ in high- \citep{2005ApJ...632L..95F, 2013A&A...554A.100I, 2013A&A...554A..78P, 2015A&A...581A..71F, 2016MNRAS.455.1428R, 2017ApJ...849..139Z, 2017MNRAS.467.2723P, 2018ApJS..237....3S, 2019A&A...632A..57C, 2022MNRAS.512.4419P, 2023A&A...677A..15M}, intermediate- \citep{2011ApJ...743L..32P, 2014A&A...568A..65F}, and low-mass star-forming regions \citep{2011A&A...534A.100J, 2016ApJ...830L..37I, 2017A&A...597A..53L, 2020A&A...639A..87V, 2021ApJ...923..155M}.

Interferometric observations of the Orion-KL region have shown that the spatial distribution of CH$_{3}$COCH$_{3}$ is primarily concentrated in the hot core, rather than in the compact ridge, resembling that of complex cyanides \citep{2008ApJ...672..962F, 2013A&A...554A..78P}. This seems to imply a potential chemical relationship between the formation of CH$_{3}$COCH$_{3}$ and N-bearing species. Although CH$_{3}$COCH$_{3}$ has been detected in various astronomical sources, its study has largely been confined to individual targets observed with interferometers or to small samples observed using single-dish telescopes. Hence, its physical and chemical properties remain poorly understood. Given that COMs are typically localized in compact regions within star-forming environments \citep{2017A&A...597A..53L, 2022MNRAS.511.3463Q}, comprehensive interferometric studies targeting CH$_{3}$COCH$_{3}$ and chemically associated molecules across a larger sample with a similar angular resolution and spectral setup are essential.

We conducted observations of 11 massive protostellar clumps using Band 7 of the Atacama Large Millimeter/submillimeter Array (ALMA). These data serve as a pilot project for the ALMA Three-millimeter Observations of Massive Star-Forming Regions (ATOMS) survey and are a subsample of ATOMS \citep{2020MNRAS.496.2790L, 2024ApJS..270....9X}. The 11 clumps have masses ranging from (1$-$22) $\times$ 10$^{3}$ M$_{\bigodot}$ and fall between the early stages of massive star formation and the H{\sc ii} region phase, with L/M ratios ranging from 12 to 80 \citep{2024ApJS..270....9X}. Eight of the clumps enshroud ultra-compact (UC) H{\sc ii} regions, as indicated by the presence of associated H40$\alpha$ emission \citep{2022MNRAS.511.3463Q, 2023MNRAS.520.3245Z}. The primary goal of this study is to investigate the physical and chemical properties of the CH$_{3}$COCH$_{3}$ molecule by comparing it with the O-bearing molecules CH$_{3}$CHO and CH$_{3}$OH, as well as the N-bearing molecule C$_{2}$H$_{5}$CN. 

The observations and data reduction are described in Sect. \ref{sec:Observations}. Section \ref{sec:Results} presents the observational results, including the line identification, spatial distribution, and parameter calculations for CH$_{3}$COCH$_{3}$, CH$_{3}$CHO, CH$_{3}$OH, and C$_{2}$H$_{5}$CN. The chemical implications of these findings are discussed in Sect. \ref{sec:Discussion}, and the conclusions are summarized in Sect. \ref{sec:Conclusions}.
\section{Observations} \label{sec:Observations}
The sample selection for the ALMA observations is described in detail by \cite{2024ApJS..270....9X} and \cite{2024ApJ...962...13C}. The ALMA Band-7 data for 11 massive protoclusters were obtained between May 18 and May 20, 2018 (Project ID: 2017.1.00545.S; PI: Tie Liu), using 43 12-meter antennas in the C43-1 configuration, with baselines ranging from 15.0 to 313.7 meters. The Band-7 observations included four spectral windows (SPWs): 342.36$-$344.24 GHz (SPW 31), 344.25$-$346.09 GHz (SPW 29), 356.60$-$357.07 GHz (SPW 27), and 354.27$-$354.74 GHz (SPW 25). SPWs 27 and 25 provided a bandwidth of 469 MHz with a spectral resolution of 0.24 km s$^{-1}$, while SPWs 31 and 29 offered a bandwidth of 1875 MHz with a spectral resolution of 0.98 km s$^{-1}$. In this paper we present spectral line data from the broader SPWs, SPW 31 and SPW 29. The calibrators used for flux and bandpass calibration were J1427-4206, J1517-2422, and J1924-2914, while phase calibration was performed using J1524-5903, J1650-5044, and J1733-3722. Data calibration was carried out using Common Astronomy Software Applications (CASA; \citealt{2007ASPC..376..127M}) version 5.1.15. The deconvolution was performed using the ``hogbom'' algorithm with a weighting parameter of ``briggs,'' and the robust parameter was set to 0.5 to balance sensitivity and resolution. Continuum imaging and line cubes were generated with the TCLEAN task in CASA 5.3. The continuum images were constructed from line-free channels across four SPWs. Due to the rich spectral line features of the sample, fewer than 10\% of the channels were available as line-free, resulting in relatively poor continuum sensitivity. Self-calibration was applied to improve the quality of the continuum images. The self-calibration solutions derived from the continuum images were applied to the line cubes. Primary beam correction was applied using pblimit = 0.2. The final synthesized beam sizes for both the continuum images and line cubes range from 0.7 to 1.0 arcsec. The average sensitivity of the continuum images is $\sim$ 1.2 mJy beam$^{-1}$, and the mean sensitivity of the line cubes is $\sim$ 4.7 mJy beam$^{-1}$ per channel.

\section{Results} \label{sec:Results}
\subsection{Line identifications} \label{subsec:li}

\begin{figure*}[h!]
        \centering
        \includegraphics[width=0.9\linewidth]{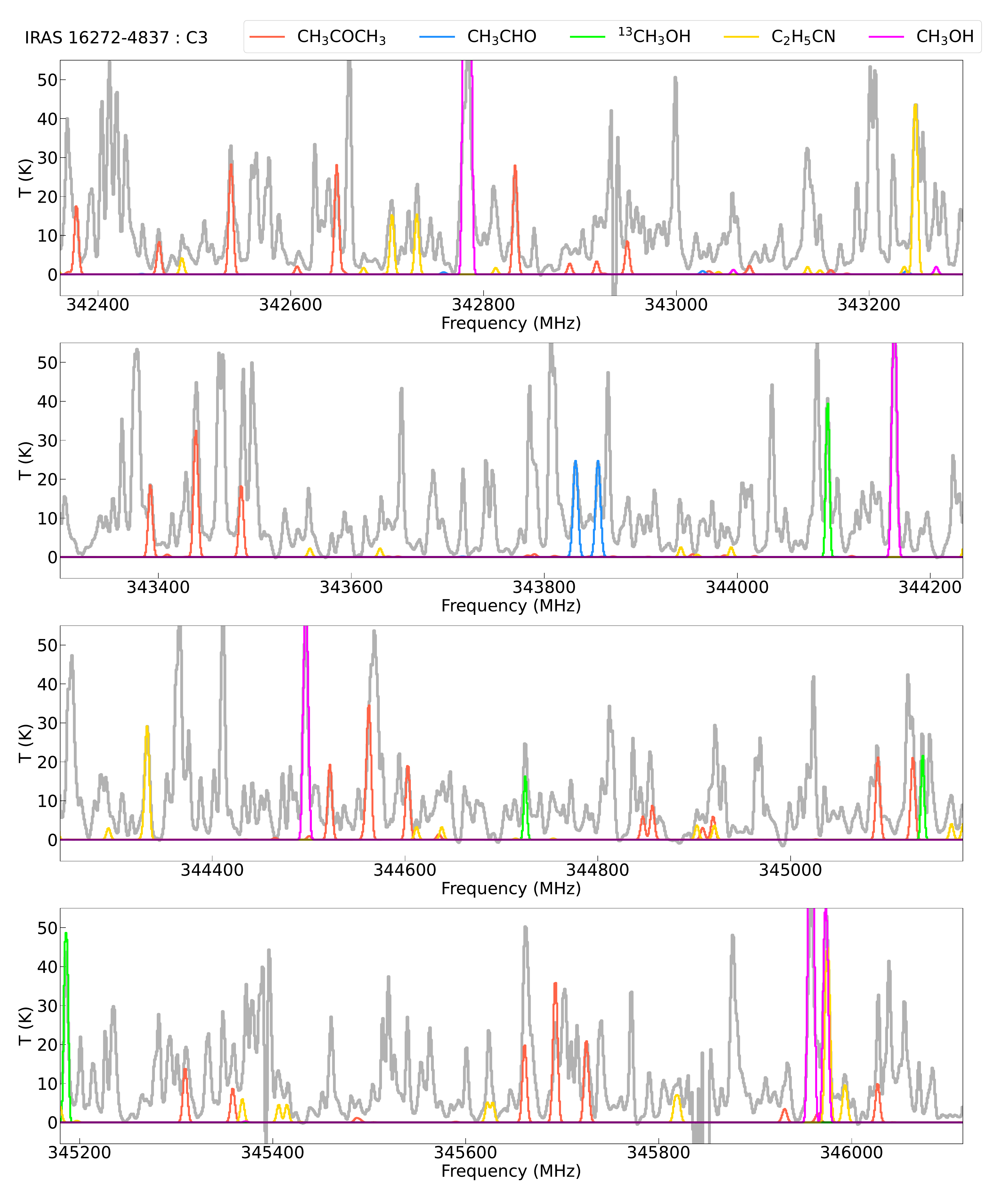}
        \caption{Spectra of IRAS 16272-4837 C3 in the frequency range 342.4$-$346.1 GHz. The observed spectra are shown as gray curves, and the XCLASS-modeled spectra as colored curves. Spectra of other sources are presented in Fig. \ref{fig5}.} \label{fig1}
\end{figure*}

\begin{figure*}[h!]
        \centering
        \includegraphics[width=\linewidth]{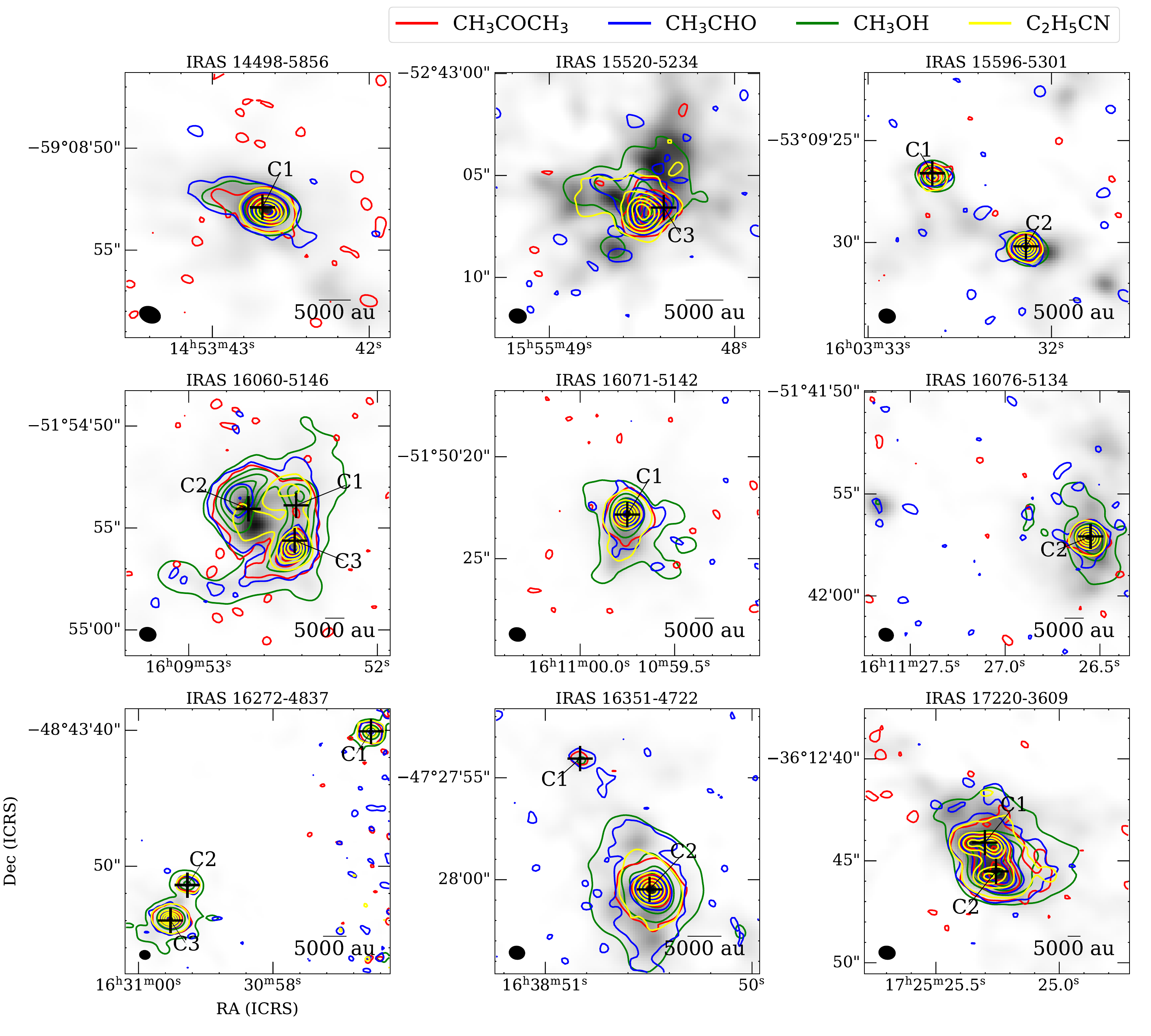}
        \caption{Continuum emission in gray scale at 870 $\mu$m overlaid with the integrated intensities of CH$_{3}$COCH$_{3}$, CH$_{3}$CHO, CH$_{3}$OH, and C$_{2}$H$_{5}$CN in nine high-mass star-forming regions. Colored contours represent the integrated intensities of each molecule. Contour levels are stepped by 10\% of the peak values, with the outermost levels as follows: 4\% for I14498, 6\% for I15520, 10\% for I15596, 1.3\% for I16060, 1.3\% for I16071, 5\% for I16076, 1\% for I16272, 0.8\% for I16351, and 1.8\% for I17220. Line-rich cores are named according to \cite{2024MNRAS.533.1583L}. The cross symbols mark the peak positions of the continuum emissions. The synthesized beam sizes are shown in the lower left, and scale bars are in the lower right.} \label{fig2}
\end{figure*}

A total of 145 dense cores were identified across the 11 clumps, and 28 cores from 9 clumps were classified as line-rich \citep{2024ApJ...962...13C}. Spectra were extracted from a single pixel at the continuum peak position of the dense core and averaged over the beam. In this work, we used the eXtended CASA Line Analysis Software Suite (XCLASS\footnote{\url{https://xclass.astro.uni-koeln.de}}; \citealt{2017A&A...598A...7M}) to identify spectral lines in the 28 line-rich cores. XCLASS accesses two molecular databases, the Cologne Database for Molecular Spectroscopy (CDMS\footnote{\url{http://cdms.de}}; \citealt{2001A&A...370L..49M, 2005JMoSt.742..215M}) and the Jet Propulsion Laboratory (JPL\footnote{\url{http:///spec.jpl.nasa.gov}}; \citealt{1998JQSRT..60..883P}) catalogs. Under the assumption of local thermodynamic equilibrium (LTE), XCLASS solves the radiative transfer equation and generates synthetic spectra for specific molecular transitions, accounting for factors such as source size, beam filling factor, line profile, opacity, and blending. The primary parameters for fitting each molecule include source size ($\theta$), rotation temperature (T), column density (N), line width ($\Delta$V), and velocity offset (V$_{\rm off}$). In our study, the source size was taken as the deconvolved size of the continuum emission (see Table \ref{tab1}), which was derived from 2D Gaussian fitting in CASA \citep{2024ApJ...962...13C}. As shown in Fig. \ref{fig2}, in some sources, the molecular line emissions appear more extended than the continuum emissions. This may lead to an underestimation of the source size, resulting in an overestimate of the fitted column densities. XCLASS also integrates with the model optimizer package, Modeling and Analysis Generic Interface for eXternal numerical codes (MAGIX; \citealt{2013A&A...549A..21M}). MAGIX optimizes the parameter solutions and provides error estimations using various algorithms. We obtained the final fitting parameters and error estimates for the molecules using the Markov chain Monte Carlo algorithm.

Transitions are considered to be detected if their line intensities exceed 3$\sigma$ noise level and the observed line intensities at different frequencies are consistent with the XCLASS model predictions. Molecular lines of CH$_{3}$COCH$_{3}$ are identified in 16 of the 28 line-rich cores. Table \ref{tab1} lists the target source names, distances, core positions, deconvolved sizes, peak flux densities, integrated flux densities, core masses, and H$_{2}$ column densities of the continuum cores, based on values reported by \cite{2024ApJ...962...13C}. The source-averaged H$_{2}$ column densities of the 16 cores are derived from the continuum flux using the following formula under the LTE assumption \citep{2008A&A...487..993K}:
\begin{equation} \label{eq1}
        \rm N (H_{2}) = \frac{S_{\nu} \eta}{\mu m_{H} \Omega \kappa_{\nu} B_{\nu}(T)},
\end{equation}
where S$_{\nu}$ is the integrated flux density, $\eta$ = 100 is the gas-to-dust mass ratio \citep{1991ApJ...380..429L,1992ApJS...82..167H}, $\mu$ = 2.8 is the mean molecular weight of the gas \citep{2008A&A...487..993K}, m$_{\rm H}$ is the mass of the hydrogen atom, $\kappa$$_{870\mu m}$ = 1.89 cm$^{2}$ g$^{-1}$ is the dust mass absorption coefficient \citep{1994A&A...291..943O}, $\Omega$ is the solid angle subtended by the source, and B$_{\nu}$(T) is the Planck function at dust temperature T (T is taken from \citealt{2024ApJ...962...13C}, and it is assumed to be equal to the gas temperature of CH$_{3}$OCHO). The 16 line-rich cores are classified into 12 high-mass line-rich cores (H, $\textgreater$8 M$_{\bigodot}$), 4 intermediate-mass line-rich cores (I, 2$-$8 M$_{\bigodot}$), and 1 low-mass line-rich core (L, $\textless$2 M$_{\bigodot}$) according to core mass. Line transitions of CH$_{3}$CHO and CH$_{3}$OH are also detected in all 16 cores, while C$_{2}$H$_{5}$CN transitions are detected in 15 cores (excluding IRAS 16060 C2). Due to the optical thickness and blending of certain CH$_{3}$OH lines (see Appendix \ref{sec:mt}), we fit the isotopolog $^{13}$CH$_{3}$OH as an alternative in the 16 cores. Figure \ref{fig1} shows sample spectra toward IRAS 16272-4837 C3, with modeled molecular spectra for CH$_{3}$COCH$_{3}$, CH$_{3}$CHO, CH$_{3}$OH, $^{13}$CH$_{3}$OH, and C$_{2}$H$_{5}$CN overlaid. Additional spectra are presented in Fig. \ref{fig5}. CH$_{3}$COCH$_{3}$, $^{13}$CH$_{3}$OH, and C$_{2}$H$_{5}$CN exhibit multiple transition lines, while CH$_{3}$CHO displays two transition lines in each core. Table \ref{tab4} lists the parameters of the CH$_{3}$COCH$_{3}$, CH$_{3}$CHO, CH$_{3}$OH, $^{13}$CH$_{3}$OH, and C$_{2}$H$_{5}$CN transitions. The upper level energy ranges of molecular transitions are 93$-$485 K for CH$_{3}$COCH$_{3}$, 227$-$459 K for CH$_{3}$OH, 36$-$144 K for $^{13}$CH$_{3}$OH, and 64$-$441 K for C$_{2}$H$_{5}$CN. CH$_{3}$CHO exhibits only two line transitions, with the same upper level energy of 166 K. We assumed that the fitting temperature of CH$_{3}$CHO is equal to the excitation temperature of CH$_{3}$COCH$_{3}$, as CH$_{3}$CHO is considered a precursor to CH$_{3}$COCH$_{3}$ \citep{2022ApJS..259....1G}. The vibrational contributions of CH$_{3}$COCH$_{3}$, CH$_{3}$CHO, and $^{13}$CH$_{3}$OH are already included in the partition function, while the partition function of C$_{2}$H$_{5}$CN only considers the contribution from the ground state in the XCLASS fitting. The vibrational correction factor (f$_{\rm vib}$) can be given as \citep{1966PASJ...18..127T}
\begin{equation} \label{eq2}
        \rm f_{vib} = \prod_{i} \frac{1}{1 - e^{-h\nu_i / kT}},
\end{equation}
where h is Planck’s constant, k is Boltzmann’s constant, $\nu$$_{i}$ is the vibrational frequency of mode i \citep{1981JMoSp..90..531H}, and T is the temperature. The column densities of C$_{2}$H$_{5}$CN have been corrected by the f$_{\rm vib}$.

\subsection{Spatial distribution \label{subsec:sd}}

We generated integrated intensity maps of CH$_{3}$COCH$_{3}$ at 342485 MHz, CH$_{3}$CHO at 343779 MHz, CH$_{3}$OH at 342730 MHz, and C$_{2}$H$_{5}$CN at 343195 MHz toward nine massive protostellar regions, as shown in Fig. \ref{fig2}. The spatial distributions of CH$_{3}$COCH$_{3}$ exhibit similarities with those of the O-bearing molecules CH$_{3}$CHO and CH$_{3}$OH, as well as the N-bearing molecule C$_{2}$H$_{5}$CN. The line emission peaks of these molecules coincide with the continuum peaks in most regions. The exceptions occur in I15520 C3, I16060 C1, C2 and C3, and I17220 C1 and C2, where the line emission peaks are offset from the continuum peaks. Intense UC H{\sc ii} regions traced by the H40$\alpha$ lines are found in I15520, I16060, and I17220 \citep{2022MNRAS.511.3463Q, 2023MNRAS.520.3245Z}, and the observed line offsets are likely influenced by UC H{\sc ii} regions. \cite{2013A&A...554A..78P} reported that the distribution of CH$_{3}$COCH$_{3}$ in Orion-KL is similar to those of N-bearing molecules, but significantly different from those of complex O-bearing molecules. However, in our larger sample, CH$_{3}$COCH$_{3}$ was found to be associated with the spatial distribution of both O- and N-bearing molecules. The resolved linear scale reported by \cite{2013A&A...554A..78P} is approximately 500 au, whereas the resolved linear scales of our samples exceed 2000 au. Future higher angular resolution observations may be able to resolve the differences in the spatial distribution of these molecules as in Orion KL. Previous studies have found that CH$_{3}$COCH$_{3}$ can only be detected in regions where C$_{2}$H$_{5}$CN is detected \citep{2008ApJ...672..962F, 2013A&A...554A..78P}. Nonetheless, I16060 C2 provides a counterexample, as C$_{2}$H$_{5}$CN is not detected in I16060 C2 (see Table \ref{tab2}). The UC H{\sc ii} region of I16060 is significantly stronger compared to I15520 and I17220 \citep{2023MNRAS.520.3245Z}, which may be linked to the absence of C$_{2}$H$_{5}$CN.

\subsection{Rotation temperatures, column densities, and abundances \label{subsec:cdrtlwama}}

The rotation temperatures of CH$_{3}$COCH$_{3}$ and C$_{2}$H$_{5}$CN, as well as the column densities of CH$_{3}$COCH$_{3}$, CH$_{3}$CHO, and C$_{2}$H$_{5}$CN in 16 line-rich cores are summarized in Table \ref{tab2}. Since two transitions of CH$_{3}$CHO have the same upper energy level, the rotation temperature was fixed to that of CH$_{3}$COCH$_{3}$ and the column density was subsequently derived. The rotation temperature of CH$_{3}$OH was assumed to be the same as that of $^{13}$CH$_{3}$OH. The column densities of CH$_{3}$OH in 15 cores were taken from \cite{2024MNRAS.533.1583L}, which were calculated from the $^{12}$C / $^{13}$C ratios based on the column densities of $^{13}$CH$_{3}$OH. The ratios of $^{12}$C / $^{13}$C follow \citep{2019ApJ...877..154Y}
\begin{equation} \label{eq3}
        \rm ^{12}C / ^{13}C = (5.08 \pm 1.10)R_{GC} + (11.86 \pm 6.60),
\end{equation}
where R$_{\rm GC}$ (in kpc) represents the distance from the Galactic Center \citep{2020MNRAS.496.2790L}. I17220 C2 is a CH$_{3}$COCH$_{3}$ core that is not compiled in \cite{2024MNRAS.533.1583L}, and the column density of CH$_{3}$OH was derived from a column density of $^{13}$CH$_{3}$OH of (2.0$\pm$0.2)$\times$10$^{17}$ cm$^{-2}$, using a $^{12}$C / $^{13}$C ratio of 18, computed by Eq. (\ref{eq3}). The rotation temperature and column density of CH$_{3}$OH are presented in Table \ref{tab3}. From Tables \ref{tab2} and \ref{tab3}, the rotation temperatures range from 92 to 163 K for CH$_{3}$COCH$_{3}$, 95 to 250 K for C$_{2}$H$_{5}$CN, and 52 to 159 K for CH$_{3}$OH. The column densities span from 10$^{15}$ to 10$^{17}$ cm$^{-2}$ for CH$_{3}$COCH$_{3}$ and C$_{2}$H$_{5}$CN, 10$^{14}$ to 10$^{16}$ cm$^{-2}$ for CH$_{3}$CHO, and 10$^{17}$ to 10$^{19}$ cm$^{-2}$ for CH$_{3}$OH. In general, the column densities of the four molecules in the H cores are 1 to 2 orders of magnitude higher than those in the I and L cores. The fractional abundance of a given molecule relative to H$_{2}$ is expressed as $\rm f_{H_2} = N / N (H_{2})$, where N represents the column density of specific molecule and N (H$_{2}$) denotes the column density of molecular hydrogen. Tables \ref{tab2} and \ref{tab3} also show the abundances of four molecules relative to H$_{2}$. The molecules CH$_{3}$COCH$_{3}$ and C$_{2}$H$_{5}$CN exhibit comparable abundances, ranging from 10$^{-9}$ to 10$^{-7}$. The abundance of CH$_{3}$CHO is found to be within the range 10$^{-9}$ to 10$^{-8}$, while CH$_{3}$OH is approximately three orders of magnitude more abundant than the other molecules, with values ranging from 10$^{-6}$ to 10$^{-5}$. The average abundances of the four molecules are (6.2$\pm$1.3)$\times$10$^{-8}$ for CH$_{3}$COCH$_{3}$, (1.6$\pm$0.4)$\times$10$^{-8}$ for CH$_{3}$CHO, (9.0$\pm$1.7)$\times$10$^{-6}$ for CH$_{3}$OH, and (7.5$\pm$1.9)$\times$10$^{-8}$ for C$_{2}$H$_{5}$CN. In clump I16351, the H core (C2) and L core (C1) residing in the same gas environment exhibit the highest and lowest CH$_{3}$COCH$_{3}$ abundances.

\begin{figure}[h!]
        \centering
        \includegraphics[width=0.9\linewidth]{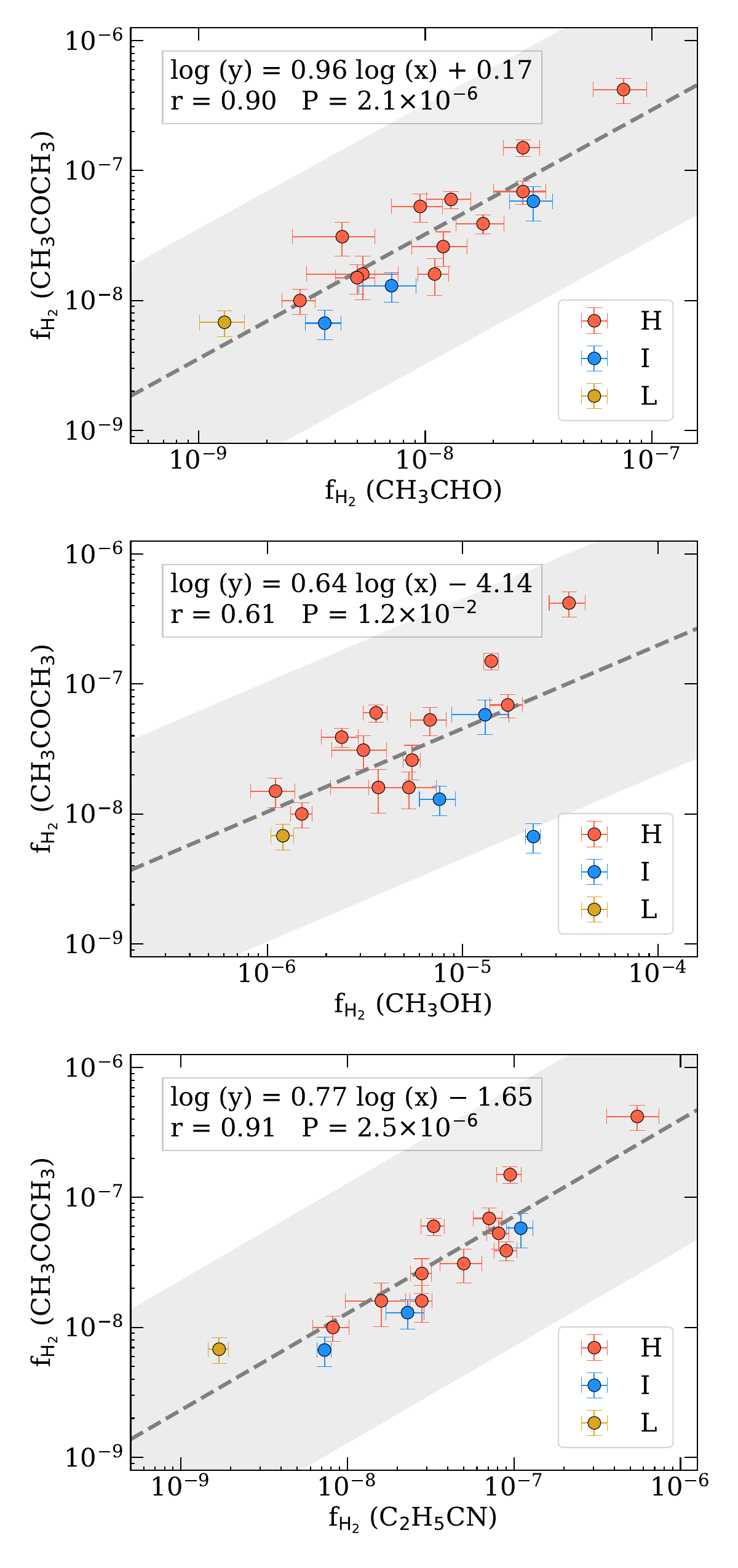}
        \caption{Abundances of CH$_{3}$COCH$_{3}$, CH$_{3}$CHO, CH$_{3}$OH, and C$_{2}$H$_{5}$CN relative to H$_{2}$. Colored circles denote different types of cores: H (high-mass line-rich cores), I (intermediate-mass line-rich cores), and L (low-mass line-rich cores). The bars represent the 1$\sigma$ errors estimated by MAGIX. Dashed lines represent linear least-squares fits to the data. The shaded regions indicate an order of magnitude scatter. Fitting results, Pearson correlation coefficients (r), and significance levels (P) are displayed in the upper left.} \label{fig3}
\end{figure}

\begin{figure}[h!]
        \centering
        \includegraphics[width=0.9\linewidth]{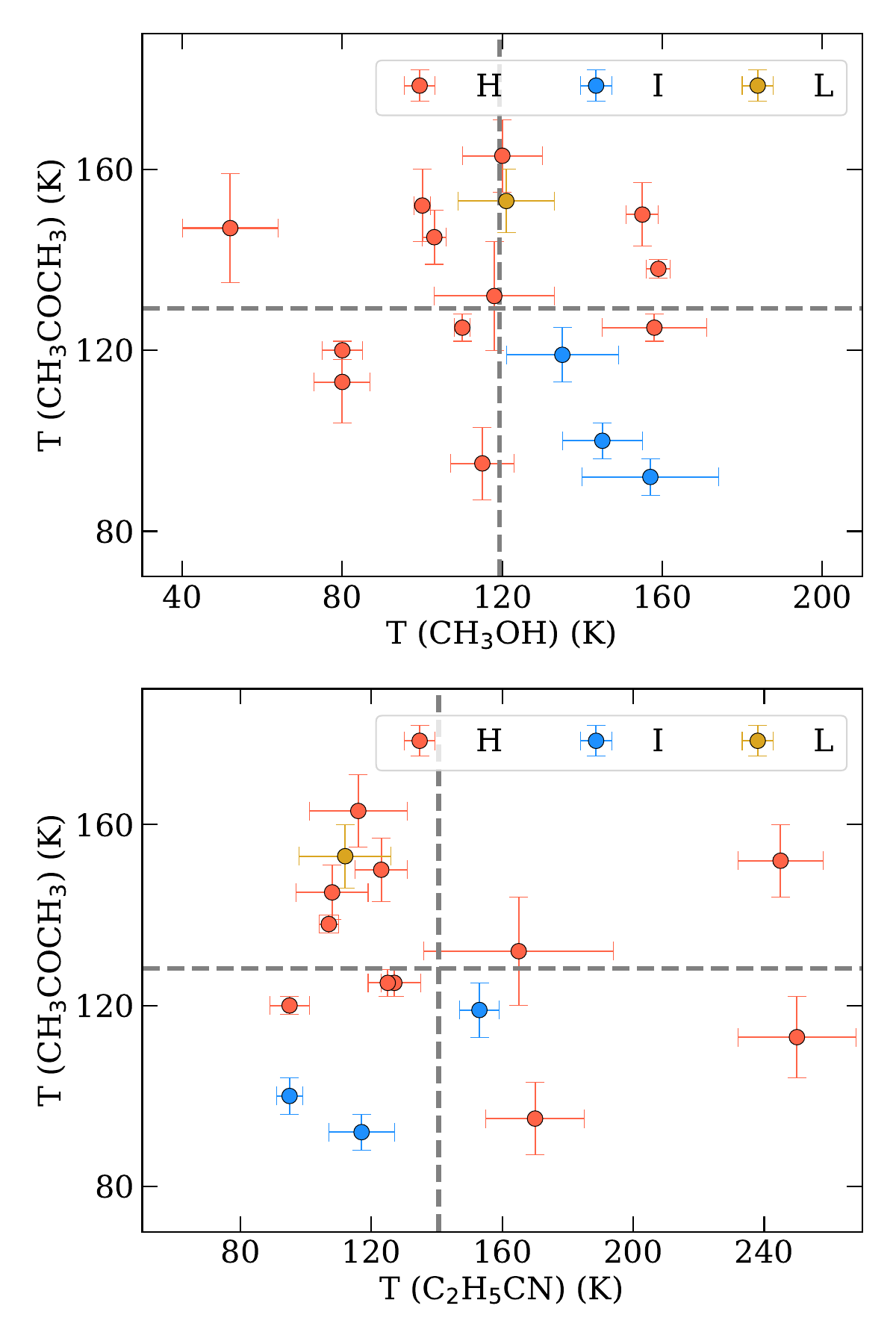}
        \caption{Comparison of the rotation temperatures of CH$_{3}$COCH$_{3}$, CH$_{3}$OH, and C$_{2}$H$_{5}$CN. Colored circles denote different core types: H (high-mass line-rich cores), I (intermediate-mass line-rich cores), and L (low-mass line-rich cores). The bars represent the 1$\sigma$ errors estimated by MAGIX. Dashed lines indicate the mean temperatures for each molecule.} \label{fig4}
\end{figure}

A comparison of the abundances of CH$_{3}$COCH$_{3}$, CH$_{3}$CHO, CH$_{3}$OH, and C$_{2}$H$_{5}$CN relative to H$_{2}$ in 16 line-rich cores is presented in Fig. \ref{fig3}. The correlation coefficients of CH$_{3}$COCH$_{3}$ with CH$_{3}$CHO, CH$_{3}$OH, and C$_{2}$H$_{5}$CN are 0.90, 0.61, and 0.91, respectively. CH$_{3}$COCH$_{3}$ exhibits a significant correlation (P\textless0.05) with CH$_{3}$OH and highly significant correlations (P\textless0.001) with CH$_{3}$CHO and C$_{2}$H$_{5}$CN. Abundance correlations between CH$_{3}$COCH$_{3}$ and C$_{2}$H$_{5}$CN, as well as between CH$_{3}$COCH$_{3}$ and CH$_{3}$OH, were also reported by \cite{2018ApJS..237....3S} in six high-mass star-forming regions, with correlation coefficients of 0.54 and 0.60, respectively. Figure \ref{fig4} further compares the rotation temperatures of CH$_{3}$COCH$_{3}$, CH$_{3}$OH, and C$_{2}$H$_{5}$CN, showing no significant correlations in temperatures among these molecules. This indicates that the abundance correlations of these molecules are not affected by temperature. The average temperatures of the three molecules are 129$\pm$6 K for CH$_{3}$COCH$_{3}$, 119$\pm$9 K for CH$_{3}$OH, and 141$\pm$11 K for C$_{2}$H$_{5}$CN. Considering the errors, the average temperatures of the three molecules are comparable.

\section{Discussion} \label{sec:Discussion}
\subsection{Comparison of the physical parameters of CH$_{3}$COCH$_{3}$ in different sources \label{subsec:atalwc}}

The rotation temperatures, column densities, and abundances of CH$_{3}$COCH$_{3}$ derived from our observations have been compared with those reported in high-, intermediate-, and low-mass star-forming regions. The mean rotation temperature of CH$_{3}$COCH$_{3}$ in the H cores from our study is 134 K, the average column density is 4.2 $\times$ 10$^{16}$ cm$^{-2}$, and the mean abundance is 7.5 $\times$ 10$^{-8}$. These values align with those found in high-mass star-forming regions \citep{2005ApJ...632L..95F, 2013A&A...554A..78P, 2016MNRAS.455.1428R, 2019A&A...632A..57C, 2022MNRAS.512.4419P}. In the I cores, the mean column density of CH$_{3}$COCH$_{3}$ is 7.9 $\times$ 10$^{15}$ cm$^{-2}$, which is comparable to that observed in intermediate-mass star-forming regions \citep{2011A&A...534A.100J, 2014A&A...568A..65F}. For the L cores, the CH$_{3}$COCH$_{3}$ column density of 1.6 $\times$ 10$^{15}$ cm$^{-2}$ lies between values reported for low-mass protostars, including IRAS 16293-2422 ($\sim$ 10$^{16}$ cm$^{-2}$; \citealt{2017A&A...597A..53L}) and IRAS 19347+0727 ($\sim$ 10$^{14}$ cm$^{-2}$; \citealt{2016ApJ...830L..37I}). These comparisons suggest that the CH$_{3}$COCH$_{3}$ emitting gas in the H, I, and L cores may have similar physical and chemical conditions as those of high-, intermediate-, and low-mass star formation regions, respectively.

\subsection{Chemistry \label{subsec:iftc}}
CH$_{3}$COCH$_{3}$, CH$_{3}$CHO, CH$_{3}$OH, and C$_{2}$H$_{5}$CN are simultaneously detected in 16 line-rich cores from our observations (except for C$_{2}$H$_{5}$CN in I16060 C2). The four molecules have also been commonly reported in other hot cores \citep{2015A&A...581A..71F, 2023A&A...677A..15M, 2014A&A...568A..65F} and hot corinos \citep{2020A&A...639A..87V}. Moreover, CH$_{3}$COCH$_{3}$, CH$_{3}$CHO, and CH$_{3}$OH have been demonstrated to be chemically linked. On the grain surface, CH$_{3}$COCH$_{3}$ and CH$_{3}$OH share a common precursor, the methyl radical (CH$_{3}$), through the following reactions \citep{2008ApJ...682..283G, 2018ApJS..237....3S}:
\begin{equation} \label{eq4}
        \rm CH_3 + OH \longrightarrow CH_3OH,
\end{equation}
\begin{equation} \label{eq5}
        \rm CH_3 + CH_3CO \longrightarrow CH_3COCH_3.
\end{equation}
Additionally, CH$_{3}$CHO acts as a precursor to CH$_{3}$COCH$_{3}$ via grain surface reactions \citep{2022ApJS..259....1G}:
\begin{equation} \label{eq6}
        \rm CH_3CHO + CH_2 \longrightarrow CH_3COCH_3,
\end{equation}
where CH$_{3}$CHO can be efficiently regenerated in the gas phase through the reaction of O with C$_{2}$H$_{5}$ \citep{2022ApJS..259....1G}. In our observations, the morphological similarities, colocated line peaks at continuum sources, and abundance correlations among CH$_{3}$COCH$_{3}$, CH$_{3}$CHO, and CH$_{3}$OH provide further evidence that these molecules are chemically related and likely form on grain surfaces. \cite{2022ApJS..259....1G} predicted the abundance of CH$_{3}$COCH$_{3}$ in the hot cores using reaction (\ref{eq6}). The model employed three warm-up timescales: 5 $\times$ 10$^{4}$ years (fast), 2 $\times$ 10$^{5}$ years (medium), and 1 $\times$ 10$^{6}$ years (slow). The predicted CH$_{3}$COCH$_{3}$ abundances for these timescales are 6.4 $\times$ 10$^{-9}$, 7.0 $\times$ 10$^{-9}$, and 1.4 $\times$ 10$^{-8}$, respectively. The average abundance of CH$_{3}$COCH$_{3}$ is (6.2$\pm$1.3)$\times$10$^{-8}$, which is comparable to the slow timescale model prediction. However, no chemical models suggest a direct association between CH$_{3}$COCH$_{3}$ and C$_{2}$H$_{5}$CN. It is possible that these two molecules coexist within similar physical environments without being chemically related.

\section{Conclusions} \label{sec:Conclusions}
We have presented a study of CH$_{3}$COCH$_{3}$ toward 11 massive protocluster clumps using ALMA Band-7 observations and analyzed the correlations among CH$_{3}$COCH$_{3}$, CH$_{3}$CHO, CH$_{3}$OH, and C$_{2}$H$_{5}$CN. Our main findings are as follows:
\begin{itemize}
        \item[1.] CH$_{3}$COCH$_{3}$, CH$_{3}$CHO, and CH$_{3}$OH lines were detected in 16 line-rich cores, while C$_{2}$H$_{5}$CN lines were detected in 15 line-rich cores, across 9 of the 11 massive protocluster clumps considered. 
        
        \item[2.] Integrated intensity maps from the nine massive protocluster clumps reveal that the spatial distribution of CH$_{3}$COCH$_{3}$ is similar to that of the O-bearing molecules CH$_{3}$CHO and CH$_{3}$OH, as well as the N-bearing molecule C$_{2}$H$_{5}$CN.
        
        \item[3.] The line peaks of the four molecules coincide with the continuum peaks in regions devoid of UC H{\sc ii} regions. In the presence of UC H{\sc ii} regions, the line peaks are offset from the continuum peaks.
        
        \item[4.] CH$_{3}$COCH$_{3}$, CH$_{3}$CHO, CH$_{3}$OH, and C$_{2}$H$_{5}$CN exhibit significant abundance correlations, and CH$_{3}$COCH$_{3}$, CH$_{3}$OH, and C$_{2}$H$_{5}$CN have similar average temperatures.
        
        \item[5.] The apparent spatial similarities and abundance correlations of CH$_{3}$COCH$_{3}$, CH$_{3}$CHO, and CH$_{3}$OH, combined with chemical models, suggest that these molecules originate from the same gas. The observed temperatures and abundances of CH$_{3}$COCH$_{3}$ are consistent with model predictions involving grain surface production.
\end{itemize}

\begin{acknowledgements}
        This paper makes use of the following ALMA data: ADS/JAO.ALMA\#2017.1.00545.S. ALMA is a partnership of ESO (representing its member states), NSF (USA), and NINS (Japan), together with NRC (Canada), MOST and ASIAA (Taiwan), and KASI (Republic of Korea), in cooperation with the Republic of Chile. The Joint ALMA Observatory is operated by ESO, AUI/NRAO, and NAOJ. This work has been supported by National Key R\&D Program of China (No. 2022YFA1603101), and by NSFC through the grants No. 12033005, No. 12073061, No. 12122307, and No. 12103045. S.-L. Qin thanks the Xinjiang Uygur Autonomous Region of China for their support through the Tianchi Program. H.-L. Liu is supported by Yunnan Fundamental Research Project (grant No. 202301AT070118). Tie Liu thanks the supports by the international partnership program of Chinese Academy of Sciences through grant No.114231KYSB20200009, and Shanghai Pujiang Program 20PJ1415500. XHL acknowledges the support by NSFC through grants no. 12473025. T.Zhang thanks the student's exchange program of the Collaborative Research Center 956, funded by the Deutsche Forschungsgemeinschaft (DFG). We would like to thank the anonymous referee for the valuable comments and suggestions, which helped improve the quality of the manuscript.
\end{acknowledgements}

\bibliographystyle{aa}
\bibliography{example}

\onecolumn
\begin{appendix} 
\section{Parameters of line-rich cores}\label{sec:polc}

\begin{table*}[h]
        \setlength{\tabcolsep}{2.5pt}
        \centering
        \small
        \caption{Physical parameters of the continuum sources.}
        \label{tab1}
        \begin{tabular}{lcccccccccc}
                \hline\hline
                \noalign{\smallskip}
                \multirow{2}{*}{Region} & \multirow{2}{*}{Core} & D & R.A. & Dec. & $\theta$$_{\rm dec}$ & I$_{\rm peak}$ & S$_{\nu}$ & M$_{\rm core}$ & N (H$_2$) & Mass $^{a}$\\
                && (kpc) & (h : m : s) & ($\circ$ : $\prime$ : {$\prime$$\prime$}) & maj($\prime$$\prime$) $\times$ min($\prime$$\prime$) & (mJy beam$^{-1}$) & (mJy) & (M$_{\bigodot}$) & (cm$^{-2}$) & Classification\\
                \noalign{\smallskip}
                \hline
                \noalign{\smallskip}
                \multirow{1}{*}{IRAS 14498-5856} & C1 & \multirow{1}{*}{3.2}
                &       14:53:42.7      &       -59:08:52.8     &       3.2$\times$1.9  &       340$\pm$22      &       2830$\pm$200    &       21.4$\pm$2.9    &       (2.1$\pm$0.2)$\times$10$^{23}$  &       H       \\
                \noalign{\smallskip}
                \hline  
                \noalign{\smallskip}
                \multirow{1}{*}{IRAS 15520-5234} & C3 & \multirow{1}{*}{2.7}
                &       15:55:48.4      &       -52:43:06.5     &       2.0$\times$1.4  &       200$\pm$5       &       1128$\pm$35     &       6.1$\pm$0.7     &       (1.8$\pm$0.1)$\times$10$^{23}$  &       I       \\
                \noalign{\smallskip}
                \hline  
                \noalign{\smallskip}
                \multirow{2}{*}{IRAS 15596-5301} & C1 & \multirow{2}{*}{10.1}
                &       16:03:32.6      &       -53:09:26.6     &       1.3$\times$0.8  &       55$\pm$4        &       160$\pm$16      &       12.3$\pm$3.3    &       (7.0$\pm$1.8)$\times$10$^{22}$  &       H       \\
                &       C2      &&      16:03:32.1      &       -53:09:30.3     &       2.2$\times$0.9  &       166$\pm$9       &       799$\pm$54      &       55.4$\pm$20.7   &       (1.7$\pm$0.6)$\times$10$^{23}$  &       H       \\
                \noalign{\smallskip}
                \hline                  
                \noalign{\smallskip}                                                                                            
                \multirow{3}{*}{IRAS 16060-5146} & C1 & \multirow{3}{*}{5.3}
                &       16:09:52.4      &       -51:54:53.8     &       2.2$\times$1.2  &       843$\pm$48      &       5000$\pm$330    &       93.6$\pm$12.7   &       (7.6$\pm$0.7)$\times$10$^{23}$  &       H       \\
                &       C2      &&      16:09:52.7      &       -51:54:53.9     &       1.6$\times$1.1  &       1530$\pm$160    &       6520$\pm$810    &       99.2$\pm$16.1   &       (1.2$\pm$0.2)$\times$10$^{24}$  &       H       \\
                &       C3      &&      16:09:52.4      &       -51:54:55.6     &       1.6$\times$1.2  &       1220$\pm$100    &       5500$\pm$550    &       105.0$\pm$15    &       (1.2$\pm$0.1)$\times$10$^{24}$  &       H       \\
                \noalign{\smallskip}
                \hline  
                \noalign{\smallskip}
                \multirow{1}{*}{IRAS 16071-5142} & C1 & \multirow{1}{*}{5.3}
                &       16:10:59.8      &       -51:50:23.1     &       1.8$\times$1.0  &       849$\pm$107     &       3840$\pm$580    &       62.9$\pm$12.4   &       (7.5$\pm$1.3)$\times$10$^{23}$  &       H       \\
                \noalign{\smallskip}
                \hline                  
                \noalign{\smallskip}                                                                                            
                \multirow{1}{*}{IRAS 16076-5134} & C2 & \multirow{1}{*}{5.3}
                &       16:11:26.5      &       -51:41:57.4     &       2.3$\times$1.5  &       160$\pm$11      &       1410$\pm$110    &       30.5$\pm$5      &       (1.9$\pm$0.2)$\times$10$^{23}$  &       H       \\
                \noalign{\smallskip}
                \hline  
                \noalign{\smallskip}
                \multirow{3}{*}{IRAS 16272-4837} & C1 & \multirow{3}{*}{2.9}
                &       16:30:57.3      &       -48:43:40.1     &       1.0$\times$0.8  &       212$\pm$8       &       549$\pm$27      &       3.3$\pm$0.4     &       (2.9$\pm$0.2)$\times$10$^{23}$  &       I       \\
                &       C2      &&      16:30:58.6      &       -48:43:51.4     &       1.2$\times$0.6  &       252$\pm$14      &       638$\pm$49      &       4.0$\pm$0.8     &       (3.9$\pm$0.7)$\times$10$^{23}$  &       I       \\
                &       C3      &&      16:30:58.8      &       -48:43:54.0     &       0.8$\times$0.8  &       1144$\pm$51     &       2560$\pm$160    &       14.0$\pm$1.7    &       (1.6$\pm$0.1)$\times$10$^{24}$  &       H       \\
                \noalign{\smallskip}
                \hline                          
                \noalign{\smallskip}                                                                                    
                \multirow{2}{*}{IRAS 16351-4722} & C1 & \multirow{2}{*}{3.0}
                &       16:38:50.8      &       -47:27:54.1     &       0.8$\times$0.6  &       113$\pm$3       &       220$\pm$9       &       1.7$\pm$0.2     &       (2.3$\pm$0.3)$\times$10$^{23}$  &       L       \\
                &       C2      &&      16:38:50.5      &       -47:28:00.8     &       2.2$\times$1.9  &       320$\pm$45      &       3010$\pm$470    &       11.6$\pm$2.3    &       (1.9$\pm$0.3)$\times$10$^{23}$  &       H       \\
                \noalign{\smallskip}
                \hline                          
                \noalign{\smallskip}                                                                                    
                \multirow{2}{*}{IRAS 17220-3609} & C1 & \multirow{2}{*}{8.0}
                &       17:25:25.3      &       -36:12:44.1     &       3.6$\times$1.4  &       554$\pm$19      &       5870$\pm$220    &       263.1$\pm$42.6  &       (4.9$\pm$0.6)$\times$10$^{23}$  &       H       \\
                &       C2      &&      17:25:25.3      &       -36:12:45.4     &       2.6$\times$1.3  &       595$\pm$26      &       4560$\pm$220    &       235.2$\pm$27.2  &       (6.6$\pm$0.4)$\times$10$^{23}$  &       H       \\
                \noalign{\smallskip}
                \hline
        \end{tabular}
        \begin{tablenotes}
                \item[ ] Notes. The basic parameters for the dense cores are adopted from \cite{2024ApJ...962...13C} and \cite{2024MNRAS.533.1583L}.
                \item[ ] $^{a}$ H = high-mass line-rich core ($\textgreater$ 8 M$_{\bigodot}$), I = intermediate-mass line-rich core (2$-$8 M$_{\bigodot}$), L = low-mass line-rich core ($\textless$ 2 M$_{\bigodot}$). 
        \end{tablenotes}
\end{table*}

\begin{sidewaystable*}
        \setlength{\tabcolsep}{5.5pt}
        \centering
        \caption{Physical parameters of CH$_{3}$COCH$_{3}$, CH$_{3}$CHO, and C$_{2}$H$_{5}$CN.}
        \label{tab2}
        \begin{tabular}{lcccccccccccc}
                \hline\hline
                \noalign{\smallskip}
                \multirow{3}{*}{Region} & \multirow{3}{*}{Core} & \multicolumn{3}{c}{CH$_{3}$COCH$_{3}$} && \multicolumn{3}{c}{CH$_{3}$CHO} && \multicolumn{3}{c}{C$_{2}$H$_{5}$CN} \\
                \noalign{\smallskip}
                \cline{3-5}\cline{7-9}\cline{11-13}
                \noalign{\smallskip}
                && T & N &      \multirow{2}{*}{f$_{\rm H_2}$} && T & N & \multirow{2}{*}{f$_{\rm H_2}$} && T & N & \multirow{2}{*}{f$_{\rm H_2}$}  \\
                && (K) & (cm$^{-2}$) &&& (K) & (cm$^{-2}$) &&& (K) & (cm$^{-2}$) &\\
                \noalign{\smallskip}
                \hline
                \noalign{\smallskip}
                \multirow{1}{*}{IRAS 14498-5856} 
                &       C1      &       152$\pm$8       &       (3.4$\pm$1.0)$\times$10$^{15}$  &       (1.6$\pm$0.5)$\times$10$^{-08}$ &&      152$\pm$8       &       (2.2$\pm$0.3)$\times$10$^{15}$  &       (1.1$\pm$0.2)$\times$10$^{-08}$ &&      245$\pm$13      &       (5.8$\pm$0.7)$\times$10$^{15}$  &       (2.8$\pm$0.4)$\times$10$^{-08}$ \\
                \noalign{\smallskip}
                \hline
                \noalign{\smallskip}
                \multirow{1}{*}{IRAS 15520-5234} 
                &       C3      &       100$\pm$4       &       (1.6$\pm$0.3)$\times$10$^{15}$  &       (8.9$\pm$1.7)$\times$10$^{-09}$ &&      100$\pm$4       &       (6.5$\pm$1.1)$\times$10$^{14}$  &       (3.6$\pm$0.6)$\times$10$^{-09}$ &&      95$\pm$4        &       (1.3$\pm$0.1)$\times$10$^{15}$  &       (7.3$\pm$0.7)$\times$10$^{-09}$ \\
                \noalign{\smallskip}
                \hline
                \noalign{\smallskip}
                \multirow{2}{*}{IRAS 15596-5301} 
                &       C1      &       113$\pm$9       &       (2.2$\pm$0.3)$\times$10$^{15}$  &       (3.1$\pm$0.9)$\times$10$^{-08}$ &&      113$\pm$10      &       (3.0$\pm$0.9)$\times$10$^{14}$  &       (4.3$\pm$1.7)$\times$10$^{-09}$ &&      250$\pm$18      &       (3.5$\pm$0.5)$\times$10$^{15}$  &       (5.0$\pm$1.4)$\times$10$^{-08}$ \\
                &       C2      &       95$\pm$8        &       (2.6$\pm$0.3)$\times$10$^{15}$  &       (1.6$\pm$0.6)$\times$10$^{-08}$ &&      95$\pm$8        &       (8.8$\pm$2.0)$\times$10$^{14}$  &       (5.3$\pm$2.3)$\times$10$^{-09}$ &&      170$\pm$15      &       (2.6$\pm$0.5)$\times$10$^{15}$  &       (1.6$\pm$0.6)$\times$10$^{-08}$ \\
                \noalign{\smallskip}
                \hline
                \noalign{\smallskip}
                \multirow{3}{*}{IRAS 16060-5146} 
                &       C1      &       120$\pm$2       &       (7.6$\pm$1.5)$\times$10$^{15}$  &       (1.0$\pm$0.2)$\times$10$^{-08}$ &&      120$\pm$2       &       (2.1$\pm$0.3)$\times$10$^{15}$  &       (2.8$\pm$0.5)$\times$10$^{-09}$ &&      95$\pm$6        &       (6.2$\pm$1.4)$\times$10$^{15}$  &       (8.2$\pm$2.0)$\times$10$^{-09}$ \\
                &       C2      &       147$\pm$12      &       (1.8$\pm$0.4)$\times$10$^{16}$  &       (1.5$\pm$0.4)$\times$10$^{-08}$ &&      147$\pm$12      &       (6.0$\pm$0.9)$\times$10$^{15}$  &       (5.0$\pm$1.0)$\times$10$^{-09}$ &&      $\cdots$        &       $\cdots$        &       $\cdots$        \\
                &       C3      &       150$\pm$7       &       (7.0$\pm$0.8)$\times$10$^{16}$  &       (6.0$\pm$0.9)$\times$10$^{-08}$ &&      150$\pm$7       &       (1.5$\pm$0.3)$\times$10$^{16}$  &       (1.3$\pm$0.3)$\times$10$^{-08}$ &&      123$\pm$8       &       (3.9$\pm$0.5)$\times$10$^{16}$  &       (3.3$\pm$0.5)$\times$10$^{-08}$ \\
                \noalign{\smallskip}
                \hline                          
                \noalign{\smallskip}                                                                                            
                \multirow{1}{*}{IRAS 16071-5142} 
                &       C1      &       138$\pm$2       &       (5.2$\pm$0.6)$\times$10$^{16}$  &       (6.9$\pm$1.4)$\times$10$^{-08}$ &&      138$\pm$2       &       (2.0$\pm$0.4)$\times$10$^{16}$  &       (2.7$\pm$0.7)$\times$10$^{-08}$ &&      107$\pm$3       &       (5.4$\pm$0.6)$\times$10$^{16}$  &       (7.1$\pm$1.4)$\times$10$^{-08}$ \\
                \noalign{\smallskip}
                \hline                          
                \noalign{\smallskip}                                                                                            
                \multirow{1}{*}{IRAS 16076-5134} 
                &       C2      &       163$\pm$8       &       (1.0$\pm$0.2)$\times$10$^{16}$  &       (5.3$\pm$1.3)$\times$10$^{-08}$ &&      163$\pm$8       &       (1.8$\pm$0.4)$\times$10$^{15}$  &       (9.5$\pm$2.4)$\times$10$^{-09}$ &&      116$\pm$15      &       (1.5$\pm$0.1)$\times$10$^{16}$  &       (8.1$\pm$1.2)$\times$10$^{-08}$ \\
                \noalign{\smallskip}
                \hline                                                  
                \noalign{\smallskip}                                                            
                \multirow{3}{*}{IRAS 16272-4837} 
                &       C1      &       119$\pm$6       &       (1.7$\pm$0.5)$\times$10$^{16}$  &       (5.8$\pm$1.7)$\times$10$^{-08}$ &&      119$\pm$6       &       (8.9$\pm$1.8)$\times$10$^{15}$  &       (3.0$\pm$0.6)$\times$10$^{-08}$ &&      153$\pm$6       &       (3.4$\pm$0.6)$\times$10$^{16}$  &       (1.1$\pm$0.2)$\times$10$^{-07}$ \\
                &       C2      &       92$\pm$4        &       (5.2$\pm$0.9)$\times$10$^{15}$  &       (1.3$\pm$0.3)$\times$10$^{-08}$ &&      92$\pm$4        &       (2.8$\pm$0.6)$\times$10$^{15}$  &       (7.1$\pm$2.0)$\times$10$^{-09}$ &&      117$\pm$10      &       (9.2$\pm$1.7)$\times$10$^{15}$  &       (2.3$\pm$0.6)$\times$10$^{-08}$ \\
                &       C3      &       145$\pm$6       &       (2.3$\pm$0.3)$\times$10$^{17}$  &       (1.5$\pm$0.2)$\times$10$^{-07}$ &&      145$\pm$6       &       (4.2$\pm$0.7)$\times$10$^{16}$  &       (2.7$\pm$0.5)$\times$10$^{-08}$ &&      108$\pm$11      &       (1.5$\pm$0.2)$\times$10$^{17}$  &       (9.5$\pm$1.6)$\times$10$^{-08}$ \\
                \noalign{\smallskip}
                \hline                          
                \noalign{\smallskip}                                                                                            
                \multirow{2}{*}{IRAS 16351-4722} 
                &       C1      &       153$\pm$7       &       (1.6$\pm$0.3)$\times$10$^{15}$  &       (6.8$\pm$1.5)$\times$10$^{-09}$ &&      153$\pm$7       &       (3.0$\pm$0.6)$\times$10$^{14}$  &       (1.3$\pm$0.3)$\times$10$^{-09}$ &&      112$\pm$14      &       (4.1$\pm$0.4)$\times$10$^{14}$  &       (1.7$\pm$0.2)$\times$10$^{-09}$ \\
                &       C2      &       132$\pm$12      &       (7.8$\pm$1.1)$\times$10$^{16}$  &       (4.2$\pm$0.9)$\times$10$^{-07}$ &&      132$\pm$12      &       (1.4$\pm$0.3)$\times$10$^{16}$  &       (7.5$\pm$2.0)$\times$10$^{-08}$ &&      165$\pm$29      &       (1.0$\pm$0.3)$\times$10$^{17}$  &       (5.5$\pm$1.9)$\times$10$^{-07}$ \\
                \noalign{\smallskip}
                \hline                          
                \noalign{\smallskip}                                                                                    
                \multirow{2}{*}{IRAS 17220-3609} 
                &       C1      &       125$\pm$3       &       (1.9$\pm$0.2)$\times$10$^{16}$  &       (3.9$\pm$0.6)$\times$10$^{-08}$ &&      125$\pm$3       &       (9.0$\pm$1.8)$\times$10$^{15}$  &       (1.8$\pm$0.4)$\times$10$^{-08}$ &&      127$\pm$8       &       (4.5$\pm$0.4)$\times$10$^{16}$  &       (9.0$\pm$1.4)$\times$10$^{-08}$ \\
                &       C2      &       125$\pm$3       &       (1.7$\pm$0.5)$\times$10$^{16}$  &       (2.6$\pm$0.8)$\times$10$^{-08}$ &&      125$\pm$4       &       (8.0$\pm$2.1)$\times$10$^{15}$  &       (1.2$\pm$0.3)$\times$10$^{-08}$ &&      125$\pm$2       &       (1.8$\pm$0.3)$\times$10$^{16}$  &       (2.8$\pm$0.4)$\times$10$^{-08}$ \\
                \noalign{\smallskip}
                \hline
        \end{tabular}
        \begin{tablenotes}
                \item[ ] Notes. The rotation temperature of CH$_{3}$CHO is assumed to be equal to that of CH$_{3}$COCH$_{3}$. ``$\cdots$'' indicates no molecular lines were detected.
        \end{tablenotes}
\end{sidewaystable*}

\begin{table*}
        \centering
        \caption{Physical parameters of CH$_{3}$OH derived from $^{13}$CH$_{3}$OH.}
        \label{tab3}
        \begin{tabular}{lcccc}
                \hline\hline
                \noalign{\smallskip}
                \multirow{3}{*}{Region} & \multirow{3}{*}{Core} & \multicolumn{3}{c}{CH$_{3}$OH} \\
                \noalign{\smallskip}
                \cline{3-5}
                \noalign{\smallskip}
                && T & N & \multirow{2}{*}{f$_{\rm H_2}$} \\
                && (K) & (cm$^{-2}$) &\\
                \noalign{\smallskip}
                \hline
                \noalign{\smallskip}
                \multirow{1}{*}{IRAS 14498-5856} 
                &       C1      &       100$\pm$2       &       (1.1$\pm$0.4)$\times$10$^{18}$ $^{a}$  &       (5.3$\pm$2.0)$\times$10$^{-06}$ \\
                \noalign{\smallskip}
                \hline
                \noalign{\smallskip}
                \multirow{1}{*}{IRAS 15520-5234} 
                &       C3      &       145$\pm$10      &       (4.1$\pm$0.3)$\times$10$^{18}$ $^{a}$  &       (2.3$\pm$2.0)$\times$10$^{-05}$ \\
                \noalign{\smallskip}
                \hline
                \noalign{\smallskip}
                \multirow{2}{*}{IRAS 15596-5301} 
                &       C1      &       80$\pm$7        &       (2.2$\pm$0.4)$\times$10$^{17}$ $^{a}$  &       (3.1$\pm$9.6)$\times$10$^{-06}$ \\
                &       C2      &       115$\pm$8       &       (6.1$\pm$1.5)$\times$10$^{17}$ $^{a}$  &       (3.7$\pm$1.6)$\times$10$^{-06}$ \\
                \noalign{\smallskip}
                \hline
                \noalign{\smallskip}
                \multirow{3}{*}{IRAS 16060-5146} 
                &       C1      &       80$\pm$5        &       (1.1$\pm$0.1)$\times$10$^{18}$ $^{a}$  &       (1.5$\pm$1.9)$\times$10$^{-06}$ \\
                &       C2      &       52$\pm$12       &       (1.3$\pm$0.3)$\times$10$^{18}$ $^{a}$  &       (1.1$\pm$2.8)$\times$10$^{-06}$ \\
                &       C3      &       155$\pm$4       &       (4.2$\pm$0.4)$\times$10$^{18}$ $^{a}$  &       (3.6$\pm$5.0)$\times$10$^{-06}$ \\
                \noalign{\smallskip}
                \hline                          
                \noalign{\smallskip}                                                                                            
                \multirow{1}{*}{IRAS 16071-5142} 
                &       C1      &       159$\pm$3       &       (1.3$\pm$0.1)$\times$10$^{19}$ $^{a}$  &       (1.7$\pm$3.2)$\times$10$^{-05}$ \\
                \noalign{\smallskip}
                \hline                          
                \noalign{\smallskip}                                                                                            
                \multirow{1}{*}{IRAS 16076-5134} 
                &       C2      &       120$\pm$10      &       (1.3$\pm$0.2)$\times$10$^{18}$ $^{a}$  &       (6.8$\pm$1.4)$\times$10$^{-06}$ \\
                \noalign{\smallskip}
                \hline                                                  
                \noalign{\smallskip}                                                            
                \multirow{3}{*}{IRAS 16272-4837} 
                &       C1      &       135$\pm$14      &       (3.7$\pm$1.2)$\times$10$^{18}$ $^{a}$  &       (1.3$\pm$4.2)$\times$10$^{-05}$ \\
                &       C2      &       157$\pm$17      &       (3.0$\pm$0.3)$\times$10$^{18}$ $^{a}$  &       (7.6$\pm$1.6)$\times$10$^{-06}$ \\
                &       C3      &       103$\pm$3       &       (2.3$\pm$0.1)$\times$10$^{19}$ $^{a}$  &       (1.4$\pm$1.2)$\times$10$^{-05}$ \\
                \noalign{\smallskip}
                \hline                          
                \noalign{\smallskip}                                                                                            
                \multirow{2}{*}{IRAS 16351-4722} 
                &       C1      &       121$\pm$12      &       (2.9$\pm$0.2)$\times$10$^{17}$ $^{a}$  &       (1.2$\pm$1.6)$\times$10$^{-06}$ \\
                &       C2      &       118$\pm$15      &       (6.6$\pm$0.8)$\times$10$^{18}$ $^{a}$  &       (3.5$\pm$7.3)$\times$10$^{-05}$ \\
                \noalign{\smallskip}
                \hline                          
                \noalign{\smallskip}                                                                                    
                \multirow{2}{*}{IRAS 17220-3609} 
                &       C1      &       110$\pm$2       &       (1.2$\pm$0.2)$\times$10$^{18}$ $^{a}$  &       (2.4$\pm$5.1)$\times$10$^{-06}$ \\
                &       C2      &       158$\pm$13      &       (3.6$\pm$0.3)$\times$10$^{18}$ $^{b}$  &       (5.5$\pm$5.6)$\times$10$^{-06}$ \\
                \noalign{\smallskip}
                \hline
        \end{tabular}
        \begin{tablenotes}
                \item[ ] Notes. The rotation temperature of CH$_{3}$OH is assumed to be equal to that of $^{13}$CH$_{3}$OH.
                \item[ ] $^{a}$ Derived by \cite{2024MNRAS.533.1583L}.
                \item[ ] $^{b}$ Derived from a column density of $^{13}$CH$_{3}$OH of (2.0$\pm$0.2)$\times$10$^{17}$ cm$^{-2}$, using a $^{12}$C / $^{13}$C ratio of 18.
        \end{tablenotes}
\end{table*}

\FloatBarrier
\clearpage

\section{Spectra of line-rich cores}\label{sec:solc}
The spectra of CH$_{3}$COCH$_{3}$, CH$_{3}$CHO, CH$_{3}$OH, $^{13}$CH$_{3}$OH, and C$_{2}$H$_{5}$CN in the frequency range 342.4$-$346.1 GHz for the 16 line-rich cores are presented in Figs. \ref{fig1} and \ref{fig5}.

\begin{figure*}[h!]
        \centering
        \includegraphics[width=0.9\linewidth]{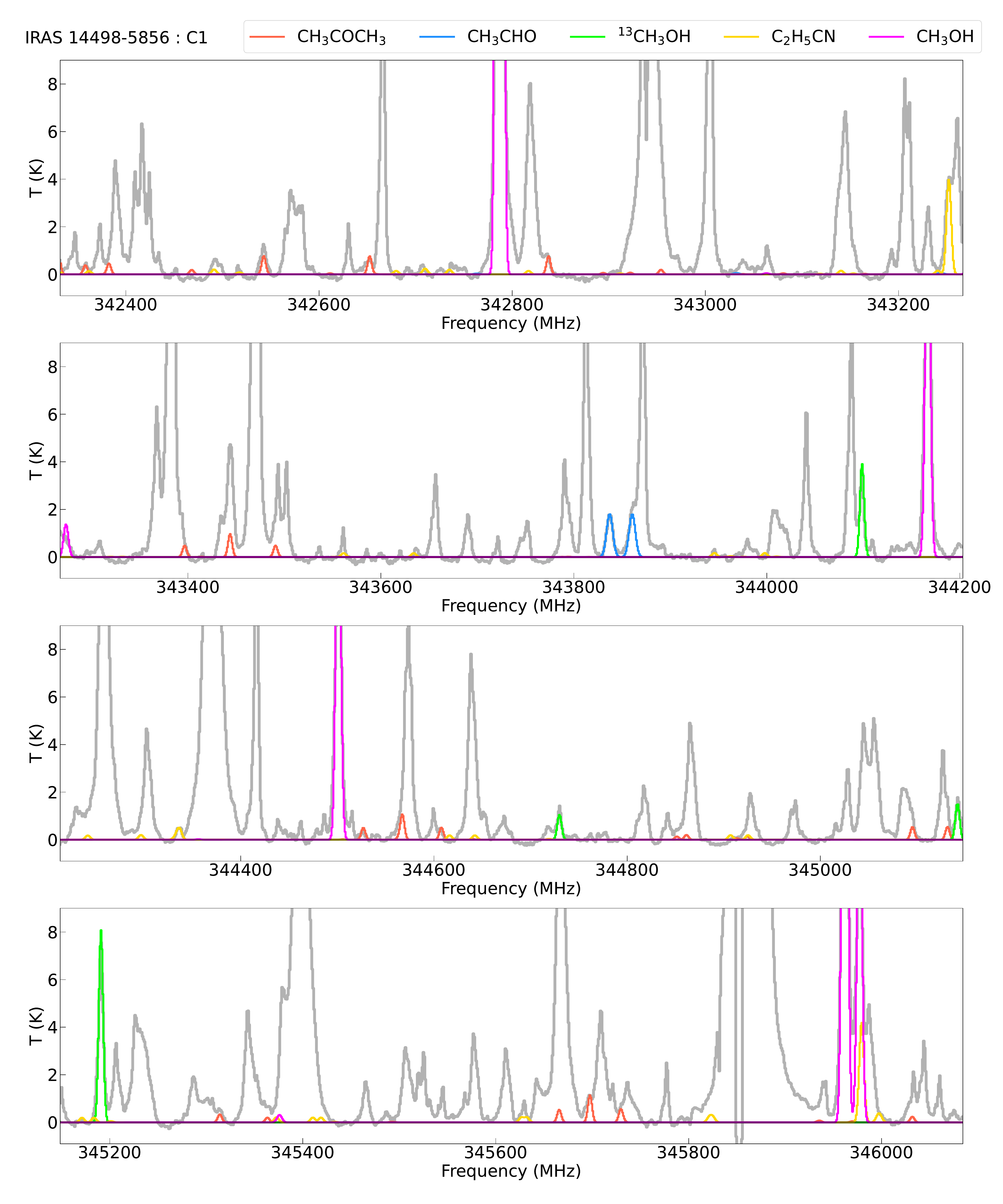}
        \caption{Same as Fig. \ref{fig1}, but for the other 15 line-rich cores.} \label{fig5}
\end{figure*} 

\addtocounter{figure}{-1}
\begin{figure*}[h!]
        \centering
        \includegraphics[width=0.9\linewidth]{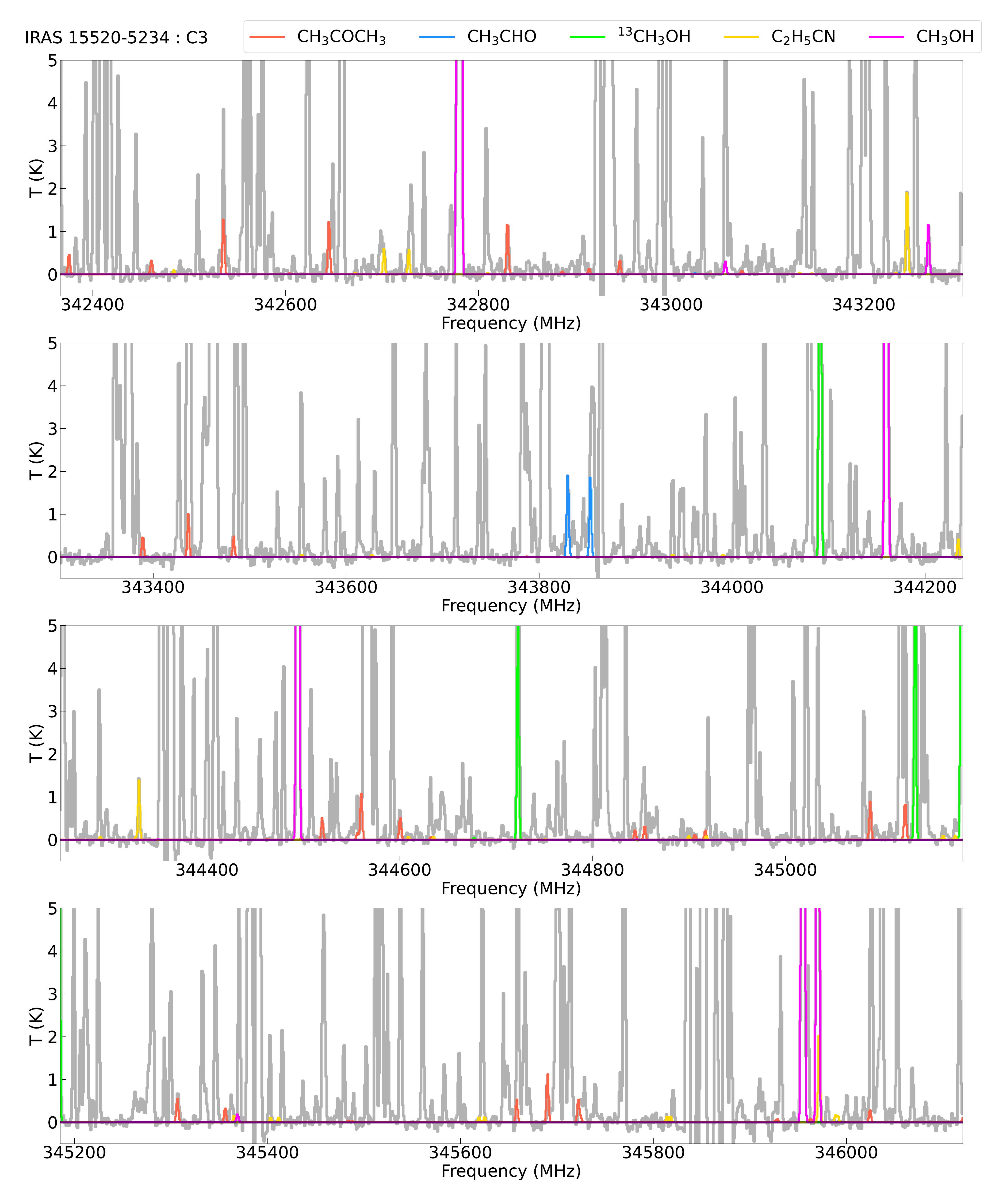}
        \caption{Continued.}
\end{figure*} 

\addtocounter{figure}{-1}
\begin{figure*}[h!]
        \centering
        \includegraphics[width=0.9\linewidth]{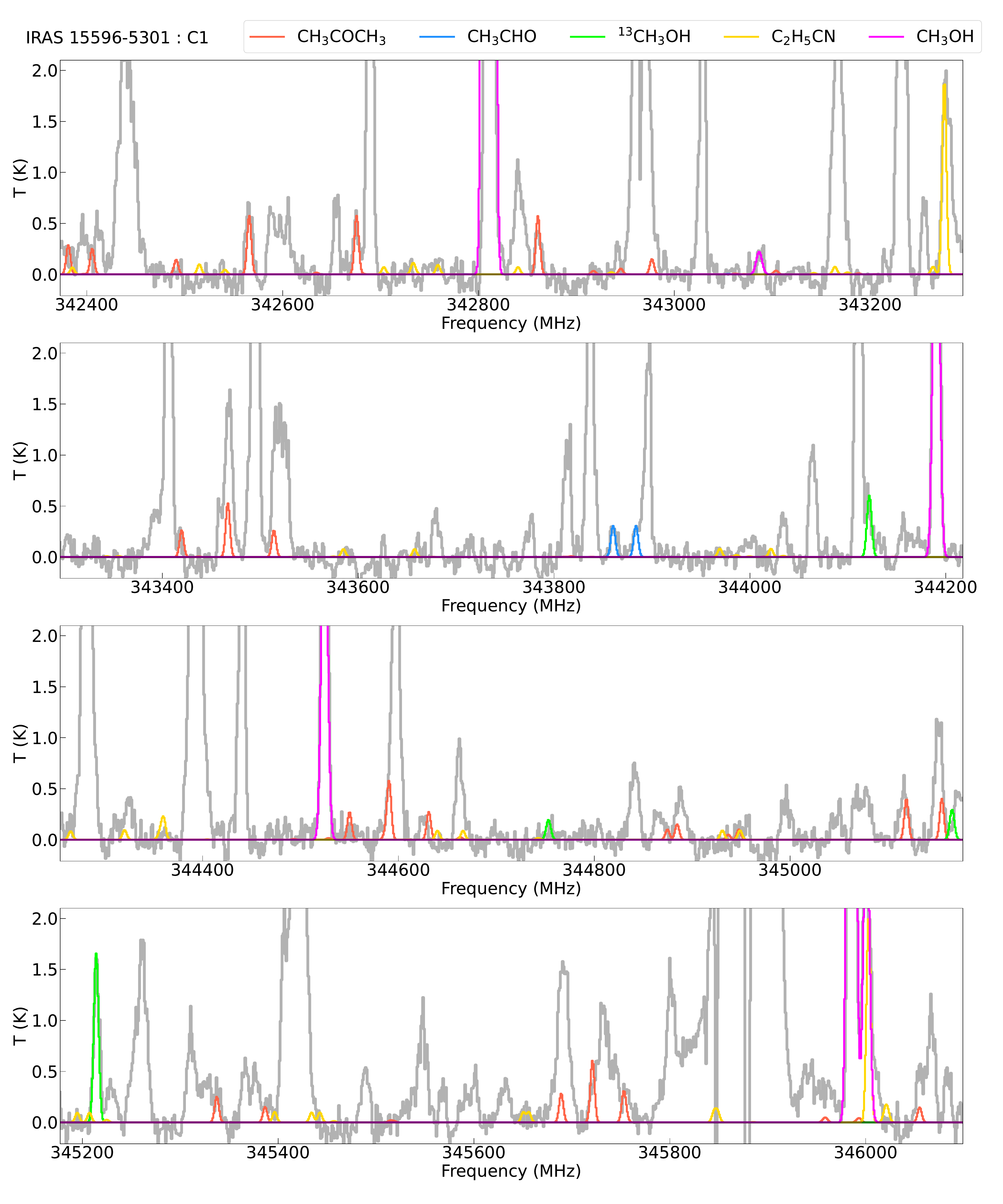}
        \caption{Continued.}
\end{figure*}

\addtocounter{figure}{-1}
\begin{figure*}[h!]
        \centering
        \includegraphics[width=0.9\linewidth]{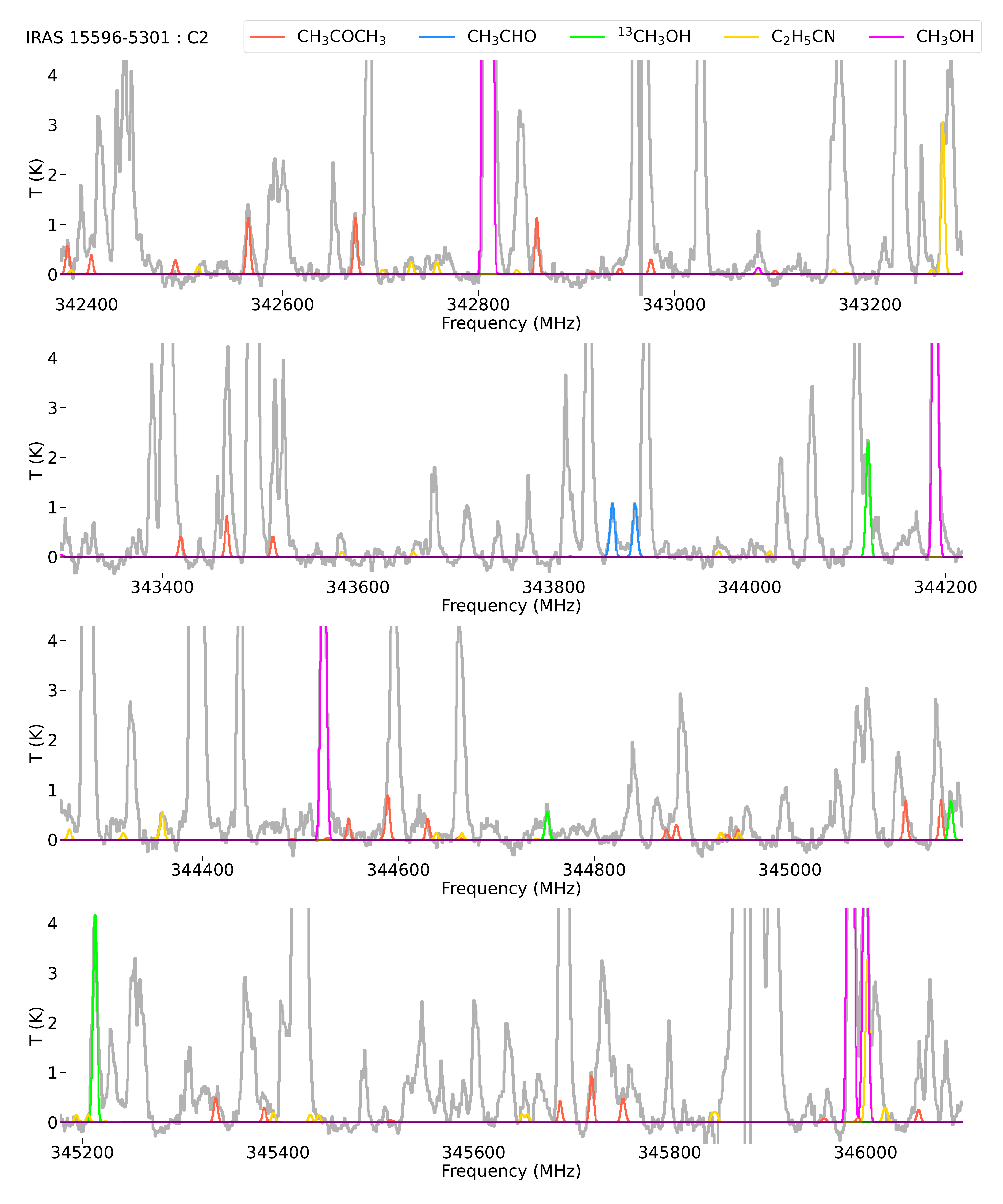}
        \caption{Continued.}
\end{figure*}

\addtocounter{figure}{-1}
\begin{figure*}[h!]
        \centering
        \includegraphics[width=0.9\linewidth]{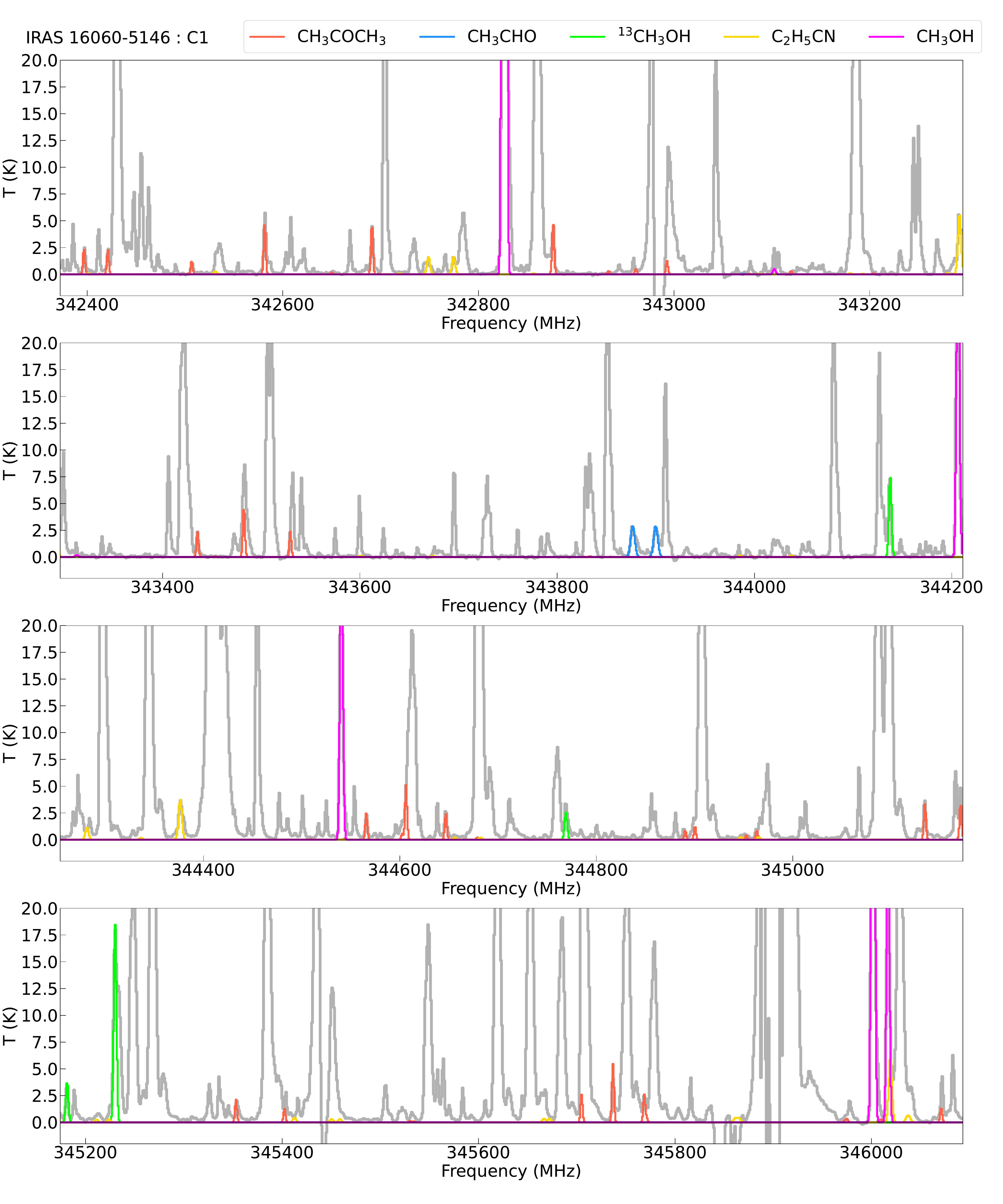}
        \caption{Continued.}
\end{figure*}

\addtocounter{figure}{-1}
\begin{figure*}[h!]
        \centering
        \includegraphics[width=0.9\linewidth]{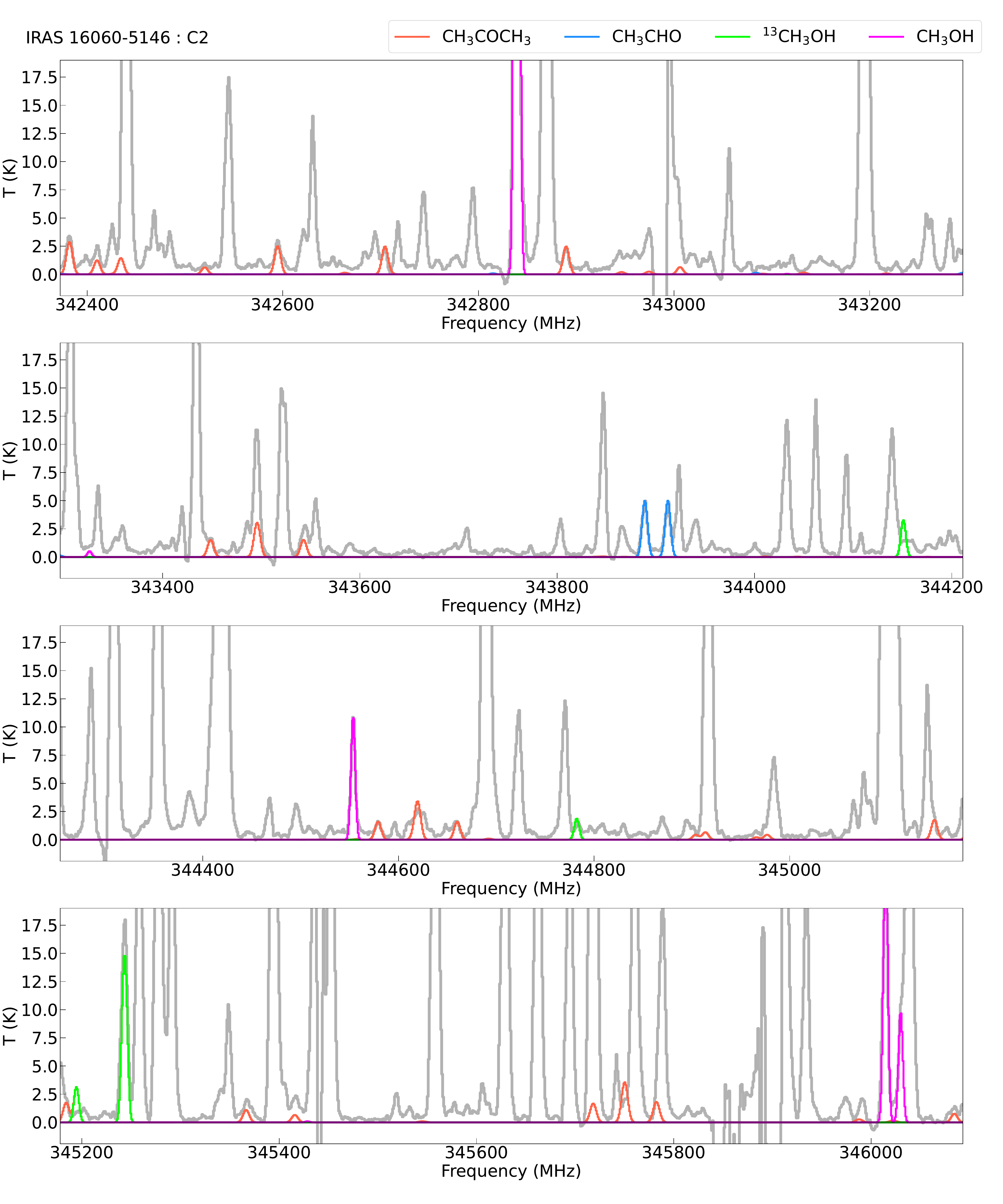}
        \caption{Continued.}
\end{figure*}

\addtocounter{figure}{-1}
\begin{figure*}[h!]
        \centering
        \includegraphics[width=0.9\linewidth]{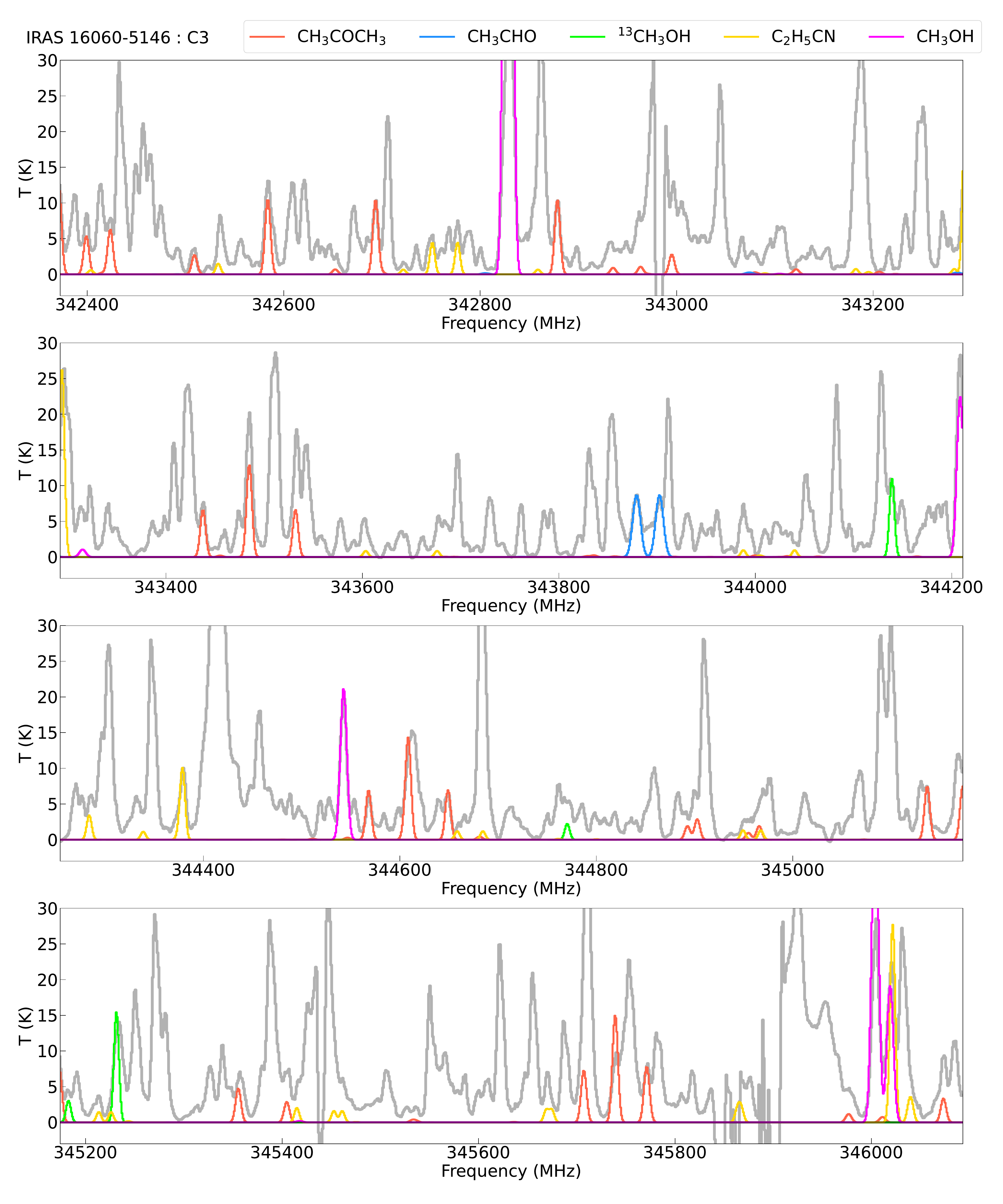}
        \caption{Continued.}
\end{figure*}

\addtocounter{figure}{-1}
\begin{figure*}[h!]
        \centering
        \includegraphics[width=0.9\linewidth]{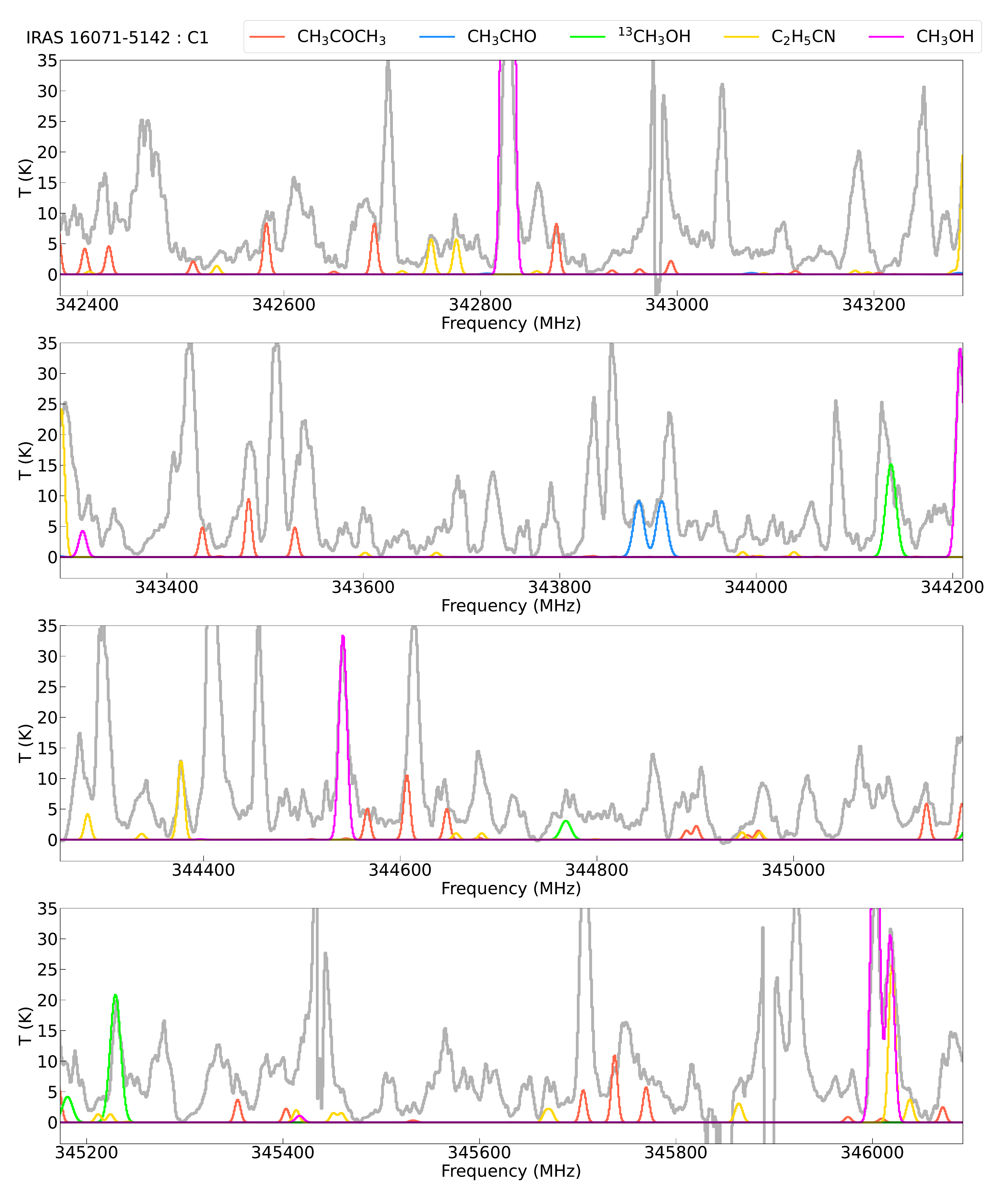}
        \caption{Continued.}
\end{figure*}

\addtocounter{figure}{-1}
\begin{figure*}[h!]
        \centering
        \includegraphics[width=0.9\linewidth]{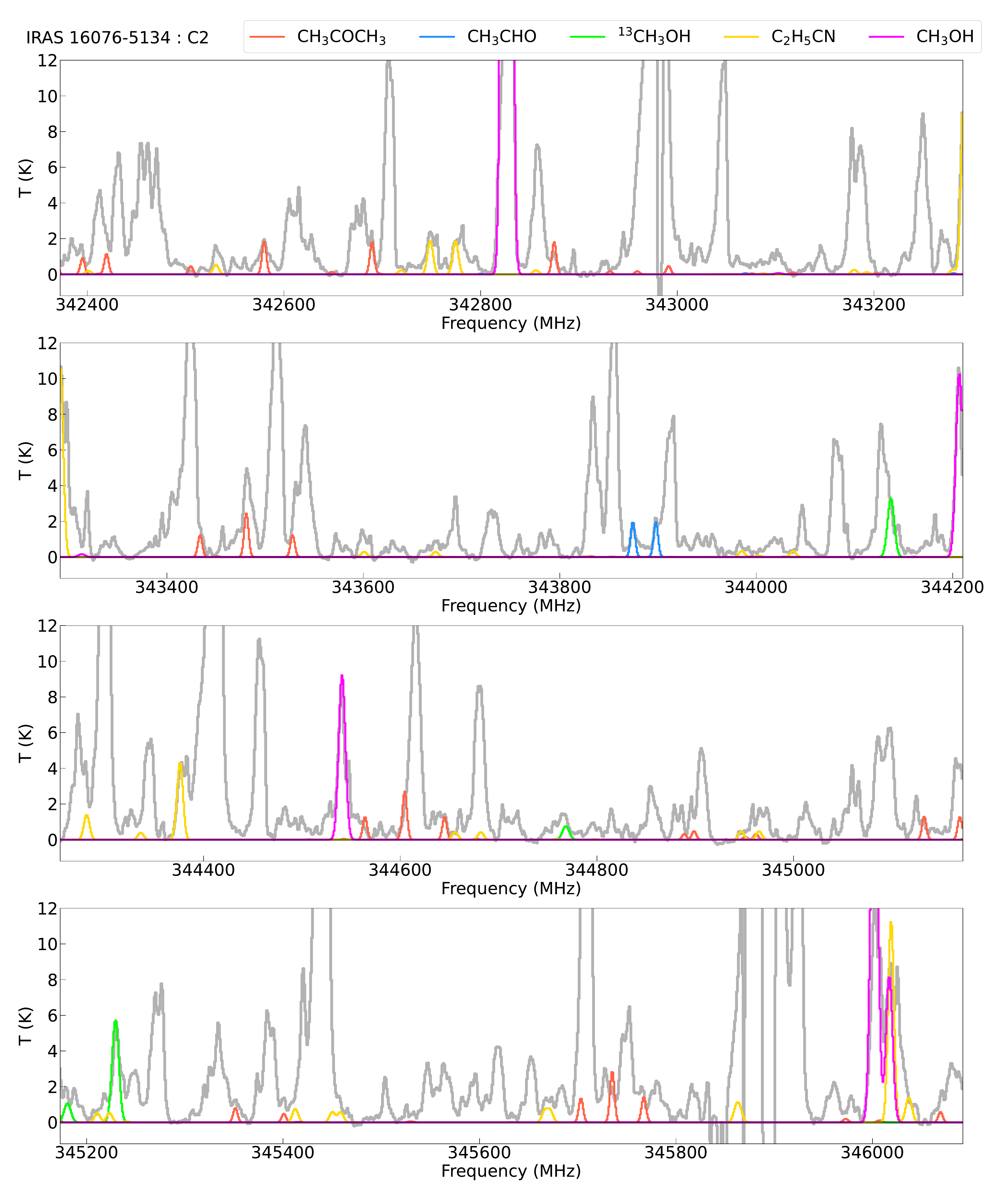}
        \caption{Continued.}
\end{figure*}

\addtocounter{figure}{-1}
\begin{figure*}[h!]
        \centering
        \includegraphics[width=0.9\linewidth]{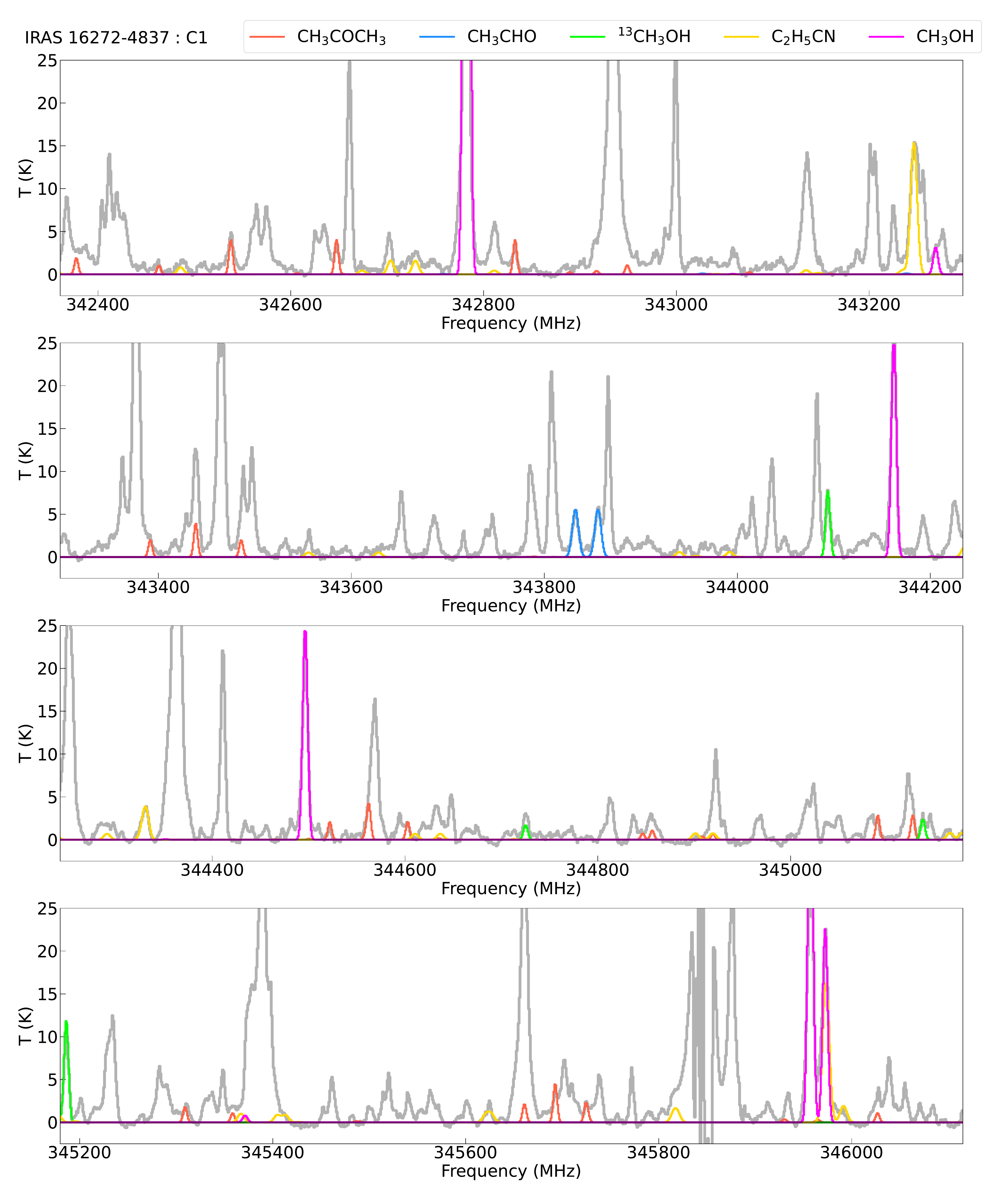}
        \caption{Continued.}
\end{figure*}

\addtocounter{figure}{-1}
\begin{figure*}[h!]
        \centering
        \includegraphics[width=0.9\linewidth]{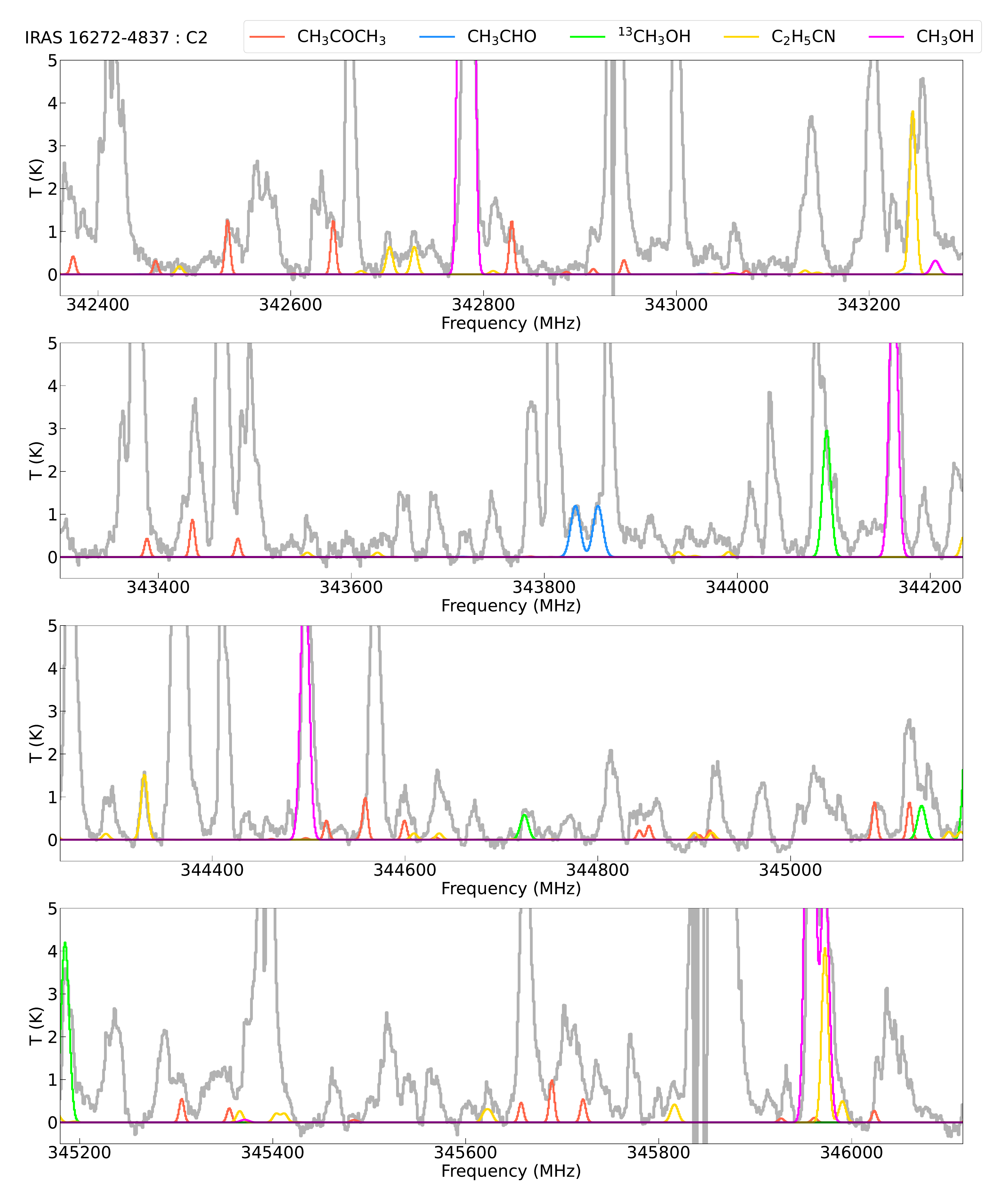}
        \caption{Continued.}
\end{figure*}

\addtocounter{figure}{-1}
\begin{figure*}[h!]
        \centering
        \includegraphics[width=0.9\linewidth]{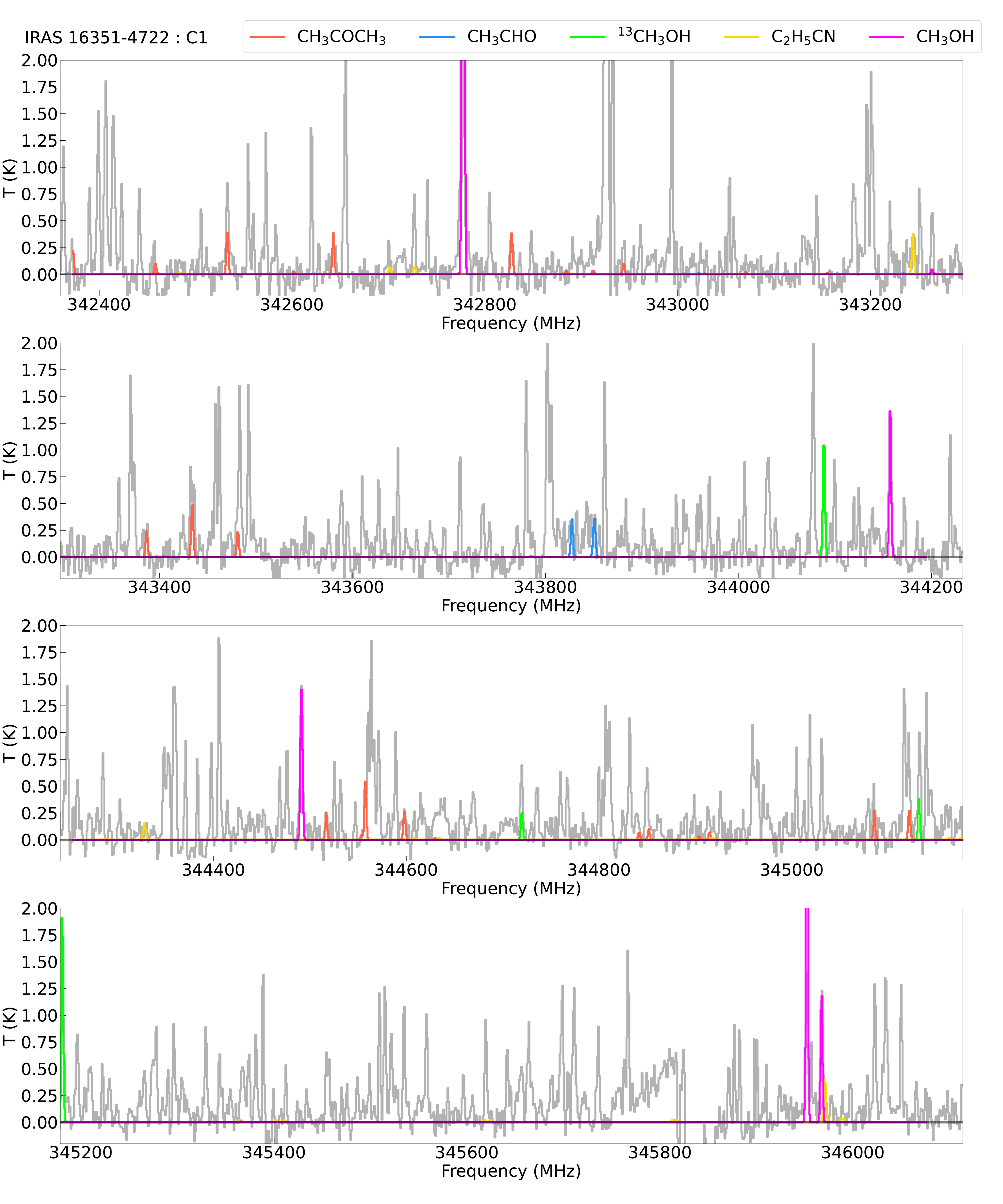}
        \caption{Continued.}
\end{figure*}

\addtocounter{figure}{-1}
\begin{figure*}[h!]
        \centering
        \includegraphics[width=0.9\linewidth]{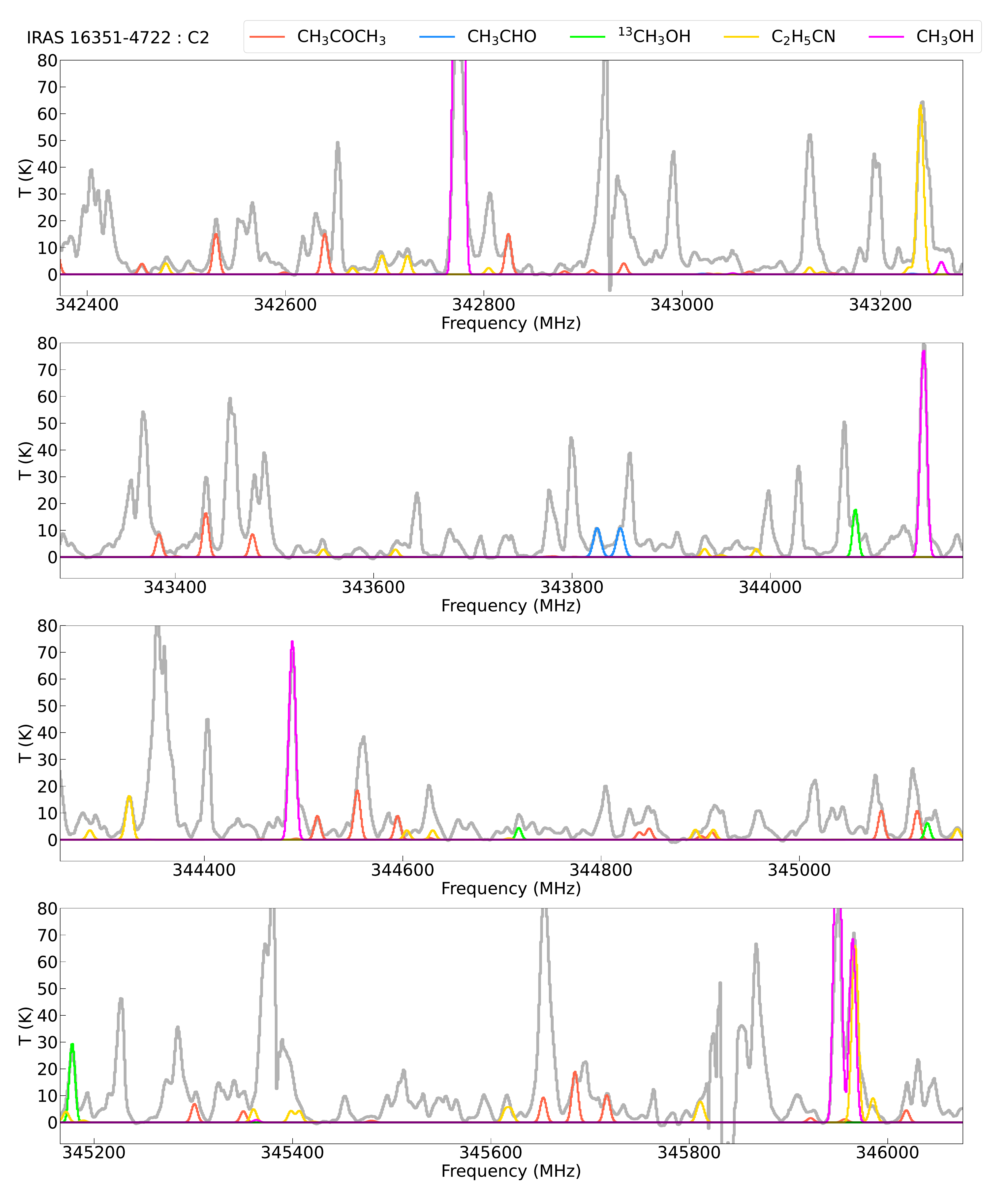}
        \caption{Continued.}
\end{figure*}

\addtocounter{figure}{-1}
\begin{figure*}[h!]
        \centering
        \includegraphics[width=0.9\linewidth]{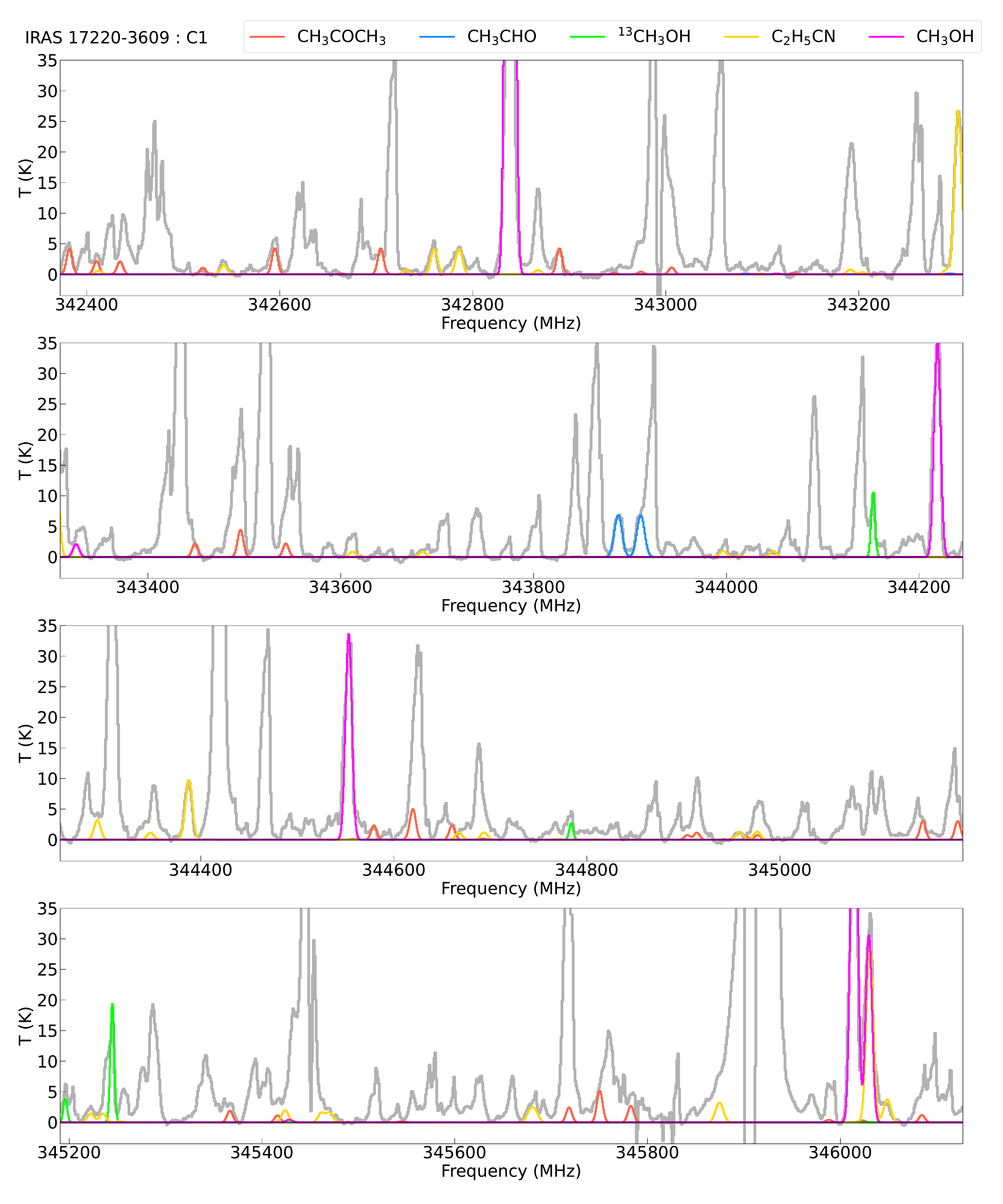}
        \caption{Continued.}
\end{figure*}

\addtocounter{figure}{-1}
\begin{figure*}[h!]
        \centering
        \includegraphics[width=0.9\linewidth]{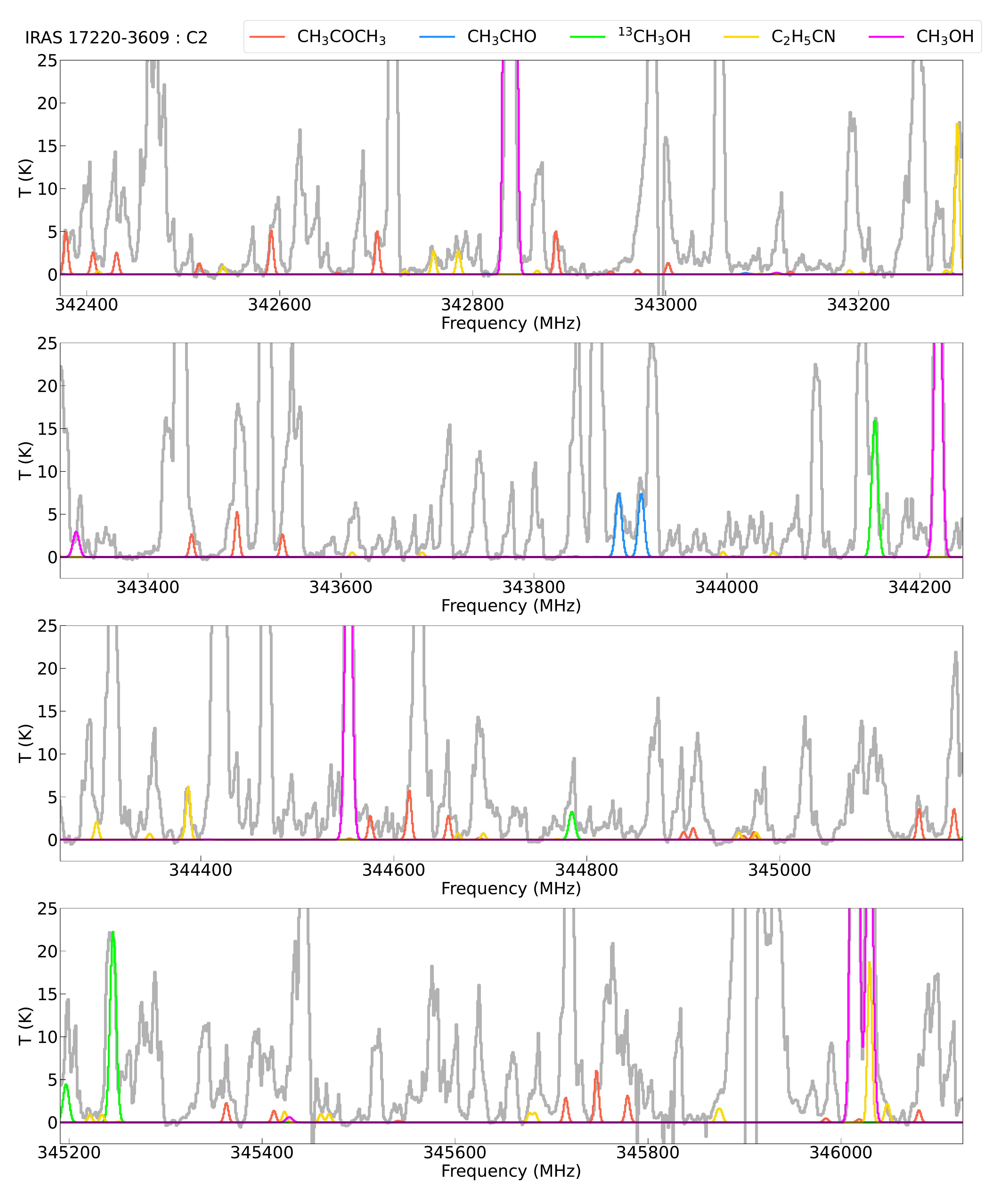}
        \caption{Continued.}
\end{figure*}

\FloatBarrier

\section{Molecular transitions}\label{sec:mt}
The rotational transitions of the molecules detected in 16 line-rich cores are presented in Table \ref{tab4}. Molecular parameters for CH$_{3}$COCH$_{3}$ and CH$_{3}$CHO lines are taken from the JPL catalog, while parameters for CH$_{3}$OH, $^{13}$CH$_{3}$OH, and C$_{2}$H$_{5}$CN lines are sourced from the CDMS catalog.

\begin{table}[h]
        \centering
        \small
        \caption{Detected spectral transitions of CH$_{3}$COCH$_{3}$, CH$_{3}$CHO, CH$_{3}$OH, $^{13}$CH$_{3}$OH, and C$_{2}$H$_{5}$CN.}
        \label{tab4}
        \begin{tabular}{lcccc}
                \hline\hline
                \noalign{\smallskip}
                Frequency & \multirow{2}{*}{Transition} & S$\mu$$^{2}$ & E$_{\rm up}$ & \multirow{2}{*}{Note}\\
                (MHz) && (D$^{2}$) & (K) &\\
                \hline
                \noalign{\smallskip}
                \multicolumn{5}{c}{CH$_{3}$COCH$_{3}$;v=0}                                                                                       \\
                \hline
                \noalign{\smallskip}
                342324.944      &       30(6,25)$-$29(5,24) AA  &       2018.37 &       282.73  &       1       \\
                342324.944      &       30(5,25)$-$29(6,24) AA  &       1211.16 &       282.73  &       1       \\
                342410.641      &       17(17,1)$-$16(16,1) EA  &       552.43  &       146.85  &               \\
                342485.234      &       17(17,0)$-$16(16,0) EE  &       2209.99 &       147.06  &       *       \\
                342553.854      &       42(3,39)$-$42(2,40) EE  &       456.49  &       484.93  &               \\
                342553.854      &       42(4,39)$-$42(3,40) EE  &       456.49  &       484.93  &               \\
                342594.888      &       17(17,1)$-$16(16,1) EE  &       2210.27 &       146.95  &               \\
                342780.035      &       17(17,1)$-$16(16,0) AA  &       828.92  &       147.05  &               \\
                342780.036      &       17(17,0)$-$16(16,1) AA  &       1381.37 &       147.05  &               \\
                342836.912      &       18(12,6)$-$17(11,7) AE  &       68.59   &       141.3   &       2       \\
                342864.593      &       18(12,6)$-$17(11,7) AA  &       208.58  &       141.24  &       3       \\
                342896.394      &       18(12,6)$-$17(11,7) EE  &       548.97  &       141.27  &       3       \\
                343023.152      &       18(12,6)$-$17(11,7) EA  &       133.9   &       141.31  &       2       \\
                343338.876      &       31(4,27)$-$30(5,26) EA  &       881.62  &       288.66  &       4       \\
                343338.876      &       31(5,27)$-$30(4,26) EA  &       881.62  &       288.66  &       4       \\
                343338.918      &       31(5,27)$-$30(5,26) AE  &       1322.45 &       288.66  &       4       \\
                343338.918      &       31(4,27)$-$30(4,26) AE  &       440.84  &       288.66  &       4       \\
                343386.006      &       31(4,27)$-$30(4,26) EE  &       3176.31 &       288.63  &       5       \\
                343386.006      &       31(5,27)$-$30(5,26) EE  &       3176.31 &       288.63  &       5       \\
                343386.006      &       31(4,27)$-$30(5,26) EE  &       349.64  &       288.63  &       5       \\
                343386.006      &       31(5,27)$-$30(4,26) EE  &       349.64  &       288.63  &       5       \\
                343433.081      &       31(4,27)$-$30(5,26) AA  &       2203.26 &       288.61  &       6       \\
                343433.081      &       31(5,27)$-$30(4,26) AA  &       1322.11 &       288.61  &       6       \\
                344468.948      &       32(4,29)$-$31(3,28) EA  &       955.83  &       293.77  &               \\
                344468.948      &       32(3,29)$-$31(4,28) EA  &       955.83  &       293.77  &               \\
                344468.985      &       32(3,29)$-$31(3,28) AE  &       1433.77 &       293.77  &               \\
                344468.985      &       32(4,29)$-$31(4,28) AE  &       477.95  &       293.77  &               \\
                344505.598      &       16(9,7)$-$15(8,8) EE    &       183.87  &       109.49  &       7       \\
                344509.424      &       32(4,29)$-$31(3,28) EE  &       3469.95 &       293.73  &       7       \\
                344509.424      &       32(4,29)$-$31(4,28) EE  &       353.28  &       293.73  &       7       \\
                344509.424      &       32(3,29)$-$31(3,28) EE  &       353.28  &       293.73  &       7       \\
                344509.424      &       32(3,29)$-$31(4,28) EE  &       3469.95 &       293.73  &       7       \\
                344549.874      &       32(4,29)$-$31(3,28) AA  &       2389.27 &       293.69  &               \\
                344549.874      &       32(3,29)$-$31(4,28) AA  &       1433.73 &       293.69  &               \\
                344793.57       &       18(15,3)$-$17(14,3) EA  &       397.32  &       150.91  &       8       \\
                344803.465      &       18(15,4)$-$17(14,3) AE  &       595.35  &       150.81  &       8       \\
                344855.635      &       18(15,3)$-$17(14,4) AE  &       198.44  &       150.81  &               \\
                344866.506      &       18(15,4)$-$17(14,4) EA  &       396.52  &       150.71  &       9       \\
                345037.5        &       18(15,3)$-$17(14,3) EE  &       1589.87 &       150.87  &       10      \\
                345073.724      &       18(15,4)$-$17(14,4) EE  &       1588.26 &       150.77  &       6       \\
                345256.037      &       18(15,4)$-$17(14,3) AA  &       994.36  &       150.83  &               \\
                345305.365      &       18(15,3)$-$17(14,4) AA  &       596.65  &       150.83  &       1       \\
                345607.709      &       33(2,31)$-$32(2,30) EA  &       1030.69 &       298.09  &       11      \\
                345607.709      &       33(3,31)$-$32(3,30) EA  &       1030.69 &       298.09  &       11      \\
                345607.736      &       33(3,31)$-$32(3,30) AE  &       1545.71 &       298.09  &       11      \\
                345607.736      &       33(2,31)$-$32(2,30) AE  &       515.26  &       298.09  &       11      \\
                345639.619      &       33(3,31)$-$32(3,30) EE  &       3529.95 &       298.03  &               \\
                345639.619      &       33(3,31)$-$32(2,30) EE  &       592.06  &       298.03  &               \\
                345639.619      &       33(2,31)$-$32(2,30) EE  &       3529.95 &       298.03  &               \\
                345639.619      &       33(2,31)$-$32(3,30) EE  &       592.06  &       298.03  &               \\
                345671.524      &       33(2,31)$-$32(2,30) AA  &       1545.65 &       297.98  &       1       \\
                345671.524      &       33(3,31)$-$32(3,30) AA  &       2575.79 &       297.98  &       1       \\
                345673.954      &       15(7,8)$-$14(6,9) EE    &       115.74  &       92.94   &       1       \\
                345875.586      &       24(12,12)$-$23(13,11) AE        &       152.92  &       235.24  &       12      \\
                345877.896      &       24(12,12)$-$23(13,11) EA        &       305.88  &       235.24  &       12      \\
                345912.434      &       15(7,8)$-$14(6,9) AA    &       72.39   &       92.88   &       13      \\
                345973.657      &       24(12,12)$-$23(13,11) EE        &       1221.85 &       235.24  &       6       \\
                \hline
        \end{tabular}
\end{table}

\addtocounter{table}{-1}
\begin{table*}[h]
        \centering
        \small
        \caption{Continued.}
        \begin{tabular}{lcccc}
                \hline\hline
                \noalign{\smallskip}
                Frequency & \multirow{2}{*}{Transition} & S$\mu$$^{2}$ & E$_{\rm up}$ & \multirow{2}{*}{Note}\\
                (MHz) && (D$^{2}$) & (K) &\\
                \hline
                \noalign{\smallskip}
                \multicolumn{5}{c}{CH$_{3}$CHO;v=0}                                                                                     \\
                \hline
                \noalign{\smallskip}
                343779.382      &       18(2,17)$-$17(2,16) E   &       224.63         &       166.47  &       *       \\
                343802.725      &       18(2,17)$-$17(2,16) A   &       224.54         &       166.46  &       1       \\
                \hline       
                \noalign{\smallskip}                                                            
                \multicolumn{5}{c}{CH$_{3}$OH;v=0}                                                                                      \\
                \hline
                \noalign{\smallskip}
                342729.796      &       13(1,12)$-$13(0,13) A   &       97.52         &       227.47  &       *       \\
                344109.039      &       18(2,17)$-$17(3,15) E   &       21.24         &       419.40  &       1       \\
                344443.433      &       19(1,19)$-$18(2,16) A   &       23.97         &       451.23  &               \\
                345903.916      &       16(1,15)$-$15(2,14) A   &       28.52         &       332.65  &               \\
                345919.260      &       18(3,15)$-$17(4,14) E   &       22.41         &       459.43  &       2       \\
                \hline       
                \noalign{\smallskip}                                                            
                \multicolumn{5}{c}{$^{13}$CH$_{3}$OH;v=0}                                                                               \\
                \hline
                \noalign{\smallskip}
                344040.629      &       8(3,6)$-$9(2,8) &       2.17    &       144.49         &               \\
                344671.733      &       3(3,0)$-$4(2,2) &       0.25    &       61.55         &       1       \\
                345083.793      &       2(2,0)$-$3(1,3) &       0.30    &       44.60         &               \\
                345132.599      &       4(0,4)$-$3(1,3) &       1.55    &       35.76         &               \\
                \hline
                \noalign{\smallskip}
                \multicolumn{5}{c}{C$_{2}$H$_{5}$CN;v=0}                                                                                \\
                \hline
                \noalign{\smallskip}
                342433.950      &       40(2,38)$-$39(3,37)     &       30.09         &       360.89  &       1       \\
                342651.886      &       15(5,11)$-$14(4,10)     &       9.20         &       79.39   &               \\
                342677.561      &       15(5,10)$-$14(4,11)     &       9.20         &       79.39   &               \\
                343194.574      &       38(5,33)$-$37(5,32)     &       553.54         &       347.41  &       *       \\
                343503.727      &       41(8,33)$-$41(7,34)     &       32.96         &       441.47  &       2       \\
                343576.364      &       41(8,34)$-$41(7,35)     &       32.96         &       441.47  &       3       \\
                343888.020      &       40(8,32)$-$40(7,33)     &       32.01         &       423.82  &       4       \\
                343940.188      &       40(8,33)$-$40(7,34)     &       32.01         &       423.82  &       2       \\
                344238.762      &       39(8,31)$-$39(7,32)     &       31.06         &       406.61  &       4       \\
                344275.871      &       39(8,32)$-$39(7,33)     &       31.06         &       406.61  &               \\
                344278.941      &       10(6,4)$-$9(5,5)        &       8.98         &       63.67   &               \\
                344278.941      &       10(6,5)$-$9(5,4)        &       8.98         &       63.67   &               \\
                344558.366      &       38(8,30)$-$38(7,31)     &       30.11         &       389.82  &               \\
                344584.505      &       38(8,31)$-$38(7,32)     &       30.12         &       389.82  &       2       \\
                344849.257      &       37(8,29)$-$37(7,30)     &       29.18         &       373.47  &               \\
                344867.347      &       37(8,30)$-$37(7,31)     &       29.18         &       373.47  &       5       \\
                345113.287      &       36(8,28)$-$36(7,29)     &       28.25         &       357.55  &               \\
                345125.723      &       36(8,29)$-$36(7,30)     &       28.25         &       357.55  &               \\
                345314.812      &       36(3,34)$-$35(2,33)     &       23.96         &       295.89  &       2       \\
                345352.535      &       35(8,27)$-$35(7,28)     &       27.33         &       342.06  &               \\
                345361.010      &       35(8,28)$-$35(7,29)     &       27.32         &       342.06  &               \\
                345568.673      &       34(8,26)$-$34(7,27)     &       26.40         &       327.00  &       3       \\
                345574.337      &       34(8,27)$-$34(7,28)     &       26.41         &       327.00  &       3       \\
                345763.372      &       33(8,25)$-$33(7,26)     &       25.49         &       312.37  &               \\
                345767.098      &       33(8,26)$-$33(7,27)     &       25.49         &       312.37  &               \\
                345921.198      &       39(3,37)$-$38(3,36)     &       573.72         &       344.46  &       6       \\
                345938.020      &       32(8,24)$-$32(7,25)     &       24.58         &       298.17  &               \\
                345940.515      &       32(8,25)$-$32(7,26)     &       24.58         &       298.17  &               \\
                \hline
        \end{tabular}
        \begin{tablenotes}
                \item[ ] Notes. S$\mu$$^{2}$ is the product of the line strength (S) and the square of the dipole moment ($\mu$). E$_{\rm up}$ represents the upper level energy. Asterisks mark the transitions of the molecules that are used for Fig. \ref{fig2}. 
                \item[ ] CH$_{3}$COCH$_{3}$ transitions: (1) blended with $^{34}$SO$_{2}$ transition lines; (2) blended with unidentified molecular lines; (3) blended with CS transition lines; (4) blended with H$_{2}$CS transition lines; (5) blended with H$_{2}$CCO transition lines; (6) blended with CH$_{3}$OCHO transition lines; (7) blended with CH$_{3}$OCH$_{3}$ transition lines; (8) blended with (CH$_{2}$OH)$_{2}$ transition lines; (9) blended with CH$_{2}$DOH and C$_{2}$H$_{5}$CN transition lines; (10) blended with HCOOH transition lines; (11) blended with HCCCN transition lines; (12) blended with H$_{2}$C$^{18}$O transition lines; (13) blended with C$_{2}$H$_{5}$CN transition lines.
                \item[ ] CH$_{3}$CHO transitions: (1) blended with H$_{2}$CS transition lines.
                \item[ ] CH$_{3}$OH transitions: (1) blended with CH$_{3}$OCHO (v=1) transition lines; (2) blended with C$_{2}$H$_{5}$CN transition lines. The 13(1,12)$-$13(0,13)A and 16(1,15)$-$15(2,14)A transitions of CH$_{3}$OH are optically thick.
                \item[ ] $^{13}$CH$_{3}$OH transitions: (1) blended with CH$_{3}$OCHO transition lines.
                \item[ ] C$_{2}$H$_{5}$CN transitions: (1) blended with CH$_{2}$DOH transition lines; (2) blended with unidentified molecular lines; (3) blended with CH$_{3}$OCHO (v=1) transition lines; (4) blended with CH$_{3}$OCHO transition lines; (5) blended with CH$_{2}$DOH and CH$_{3}$COCH$_{3}$ transition lines; (6) blended with CH$_{3}$OH transition lines.
        \end{tablenotes}
\end{table*}

\end{appendix}

\end{document}